\pdfoutput=1
\documentclass[twocolumn]{aastex62}
\usepackage{amsmath}
\graphicspath{{./}{figures/}}

\shorttitle{Close Companions of Quasars}
\shortauthors{Yue et al.}

\begin{document}

\title{Quasars Have Fewer Close Companions than Normal Galaxies}

\correspondingauthor{Minghao Yue}
\email{yuemh@email.arizona.edu}

\author{Minghao Yue}
\affiliation{Steward Observatory, University of Arizona,
933 N. Cherry Ave., Tucson, AZ 85721 }

\author{Xiaohui Fan}
\affiliation{Steward Observatory, University of Arizona,
933 N. Cherry Ave., Tucson, AZ 85721 }

\author{Jan-Torge Schindler}
\affiliation{Steward Observatory, University of Arizona,
933 N. Cherry Ave., Tucson, AZ 85721 }
\affiliation{Max Planck Institute for Astronomy, K\"onigstuhl 17, 69117 Heidelberg, Germany}

\author{Ian D. McGreer}
\affiliation{Steward Observatory, University of Arizona,
933 N. Cherry Ave., Tucson, AZ 85721 }

\author{Yun-Hsin Huang}
\affiliation{Steward Observatory, University of Arizona,
933 N. Cherry Ave., Tucson, AZ 85721}

\begin{abstract}

We investigate the distribution of companion galaxies around quasars
using {\em Hubble Space Telescope} ({\em HST}) Advanced Camera for Surveys 
Wide Field Camera (ACS/WFC) archival images. Our master sample
contains 532 quasars which have been observed by {\em HST} ACS/WFC,
spanning a wide range of luminosity $(-31<M_i(z=2)<-23)$ and redshift ($0.3<z<3$).
We search for companions around the quasars with 
projected distance of $10\text{ kpc}<d<100\text{ kpc}$.
PSF subtraction is performed to enhance the completeness for close companions.
The completeness is estimated to be high $(>90\%)$ even for the faintest companions of interest.
The number of physical companions is estimated by subtracting
a background density from the number density of projected companions.
We divide all the companions into three groups (faint, intermediate and bright)
according to their fluxes.
A control sample of galaxies is constructed to have similar redshift distribution 
and stellar mass range as the quasar sample using the data from {\em HST} deep fields.
We find that quasars and control sample galaxies 
have similar numbers of faint and bright companions, while
quasars show a $3.7\sigma$ deficit of intermediate companions compared to galaxies.
The numbers of companions in all three groups do  not show strong evolution with redshift, and
the number of intermediate companions around quasars decreases with quasar luminosity.
Assuming that merger-triggered quasars
have entered the final coalescence stage during which 
individual companions are no longer detectable at large separations, 
our result is consistent with a picture in which a significant fraction
of quasars is triggered by mergers.
\end{abstract}


\section{Introduction} \label{sec:intro}

Active Galactic Nuclei (AGNs) play important roles in not only the growth of supermassive
black holes (SMBHs) in the centers of galaxies, but also in the evolution
of their host galaxies. SMBHs gain most of their mass through
the AGN phase, and AGN activities can significantly influence the evolution of their host galaxies
\citep[for a recent review, see][]{kh13}.
AGN can heat up and expel gas from their host galaxies and thus quench 
star formation \citep[e.g.,][]{croton06,cicone14,spacek16}. 
%It is still unclear how AGN activity emerges in galaxies.
To draw the whole picture of SMBH and
galaxy evolution through cosmic time, it is crucial to
understand the triggering mechanism of AGNs.

Two scenarios have been proposed that can trigger AGN activities:
major mergers of galaxies (especially gas-rich ones) and secular evolution.
Major mergers can disturb gas in galaxies
and generate gas inflows that are needed to feed the central SMBH \citep[e.g.,][]{barnes91, hopkins06}.
Secular evolution happens when  instabilities
in galaxies, including those induced by galaxy bars or resulting from 
the gas inflows from the environment, 
drive gas to gradually move inward
and fuel the SMBH \citep[e.g.,][]{shlosman89,hopkins10}.

%While simulations support both of the scenarios,
While simulations have shown that both scenarios can lead to rapid SMBH growth in the AGN phase, 
observational evidence regarding which one is the dominant mechanism remains ambiguous. 
A crucial test is to measure the merging fraction of their host galaxies \citep[e.g.,][]{grodin05,karouzos14}.
The {\it{Hubble Space Telescope}} ({\it HST}) is powerful in identifying
galaxy mergers in AGNs because of its small Point Spread Function (PSF).
%, and 
%many studies have used HST images to measure the merging fraction of AGNs. 
\citet{cisternas11} analyzed the morphology of X-ray selected
AGNs; they found that their merging fraction is
the same as inactive galaxies within measurement errors.
\citet{villforth17} worked on a sample of more luminous X-ray selected AGNs 
and reached a conclusion that was similar to \citet{cisternas11}.
However, \citet{fan16} reported an enhanced merging fraction of infrared-selected AGNs.
Using imaging data from the Hyper-Supreme Camera (HSC) Survey 
\citep{aih18} which used the Subaru Telescope,
\citet{goulding17} also showed that the merging fraction of infrared-selected AGNs
is larger than that of inactive galaxies.
\citet{treister12} argued that major mergers  are only responsible to the most luminous AGNs.
These observational results, together with simulations \citep[e.g.,][]{hopkins06,hopkins10},
suggest that the two triggering mechanisms may dominate for AGNs of different properties
(e.g., luminosity, obscuration, or redshift).

The majority of previous studies used disturbed galaxy morphology
to identify recent merging events.
%{\bf you should discuss why this is difficult for quasars at high redshift, or with relatively low S/N, because of its low surface brightness etc.}
Galaxy pairs, or close companions of galaxies, is another frequently used indicator of 
galaxy mergers \citep[e.g.,][]{man12}.
%As an example of studies on the companions around AGNs,
\citet{ellison11} measured the ``pair fraction'' (i.e., the average number of close companions) 
of 11060 galaxies from Sloan Digital Sky Survey (SDSS), claiming that the merging fraction of 
emission-line-selected low-luminosity AGNs is larger than that of the galaxy control sample.

Although there have been a number of previous studies 
to constrain the merging fraction of AGNs, few of them focused on
the most luminous population of AGNs, i.e., type-1 quasars.
Unlike the low-luminosity AGNs,
quasars are usually much brighter than their host galaxies,
which makes it very difficult to detect the disturbed 
features in quasar host galaxies using ground-based imaging.
{\em HST}  imaging  can resolve some quasar host galaxies, mostly at low redshift ($z\lesssim 1$).
However, the sample sizes of studies based on {\em HST} were small,
usually containing several tens of AGNs, resulting 
%Small sample size results 
in large statistical errors. 
%which makes it hard to see any statistically significant
%differences between the merging fraction
%of quasars and inactive galaxies or other populations of AGNs.
On the other hand,  using close companions as indicators of mergers
is more accessible than disturbed host galaxy morphology for bright quasars, and can be expanded to larger samples. 
However, systematic studies on luminous quasar companions using {\em HST} are still lacking.

In this work, we measure the statistics of quasar companions and use the result
to constrain the merging fraction of quasars.
We use {\em HST} archival imaging for companion detection.
Our master sample contains 532 quasars, which is much larger 
than previous studies based on {\em HST} imaging.
%We apply PSF subtraction to improve the 
%completeness of companion detection.
The paper is organized as follows.
\S \ref{sec:data} describes the selection of 
the quasar sample and archival images.
\S \ref{sec:psfsub} describes the detection of close companions,
including the PSF subtraction method, as well as measurements of quasar companion fractions.
The selection of the galaxy control sample and the comparison between
companion fractions in quasars and normal galaxies are discussed in \S \ref{sec:control}.
 \S \ref{sec:discussion}
discusses the implication of the companion fraction in quasars in the context of merger-driver model of quasar
triggering. \S \ref{sec:sum} summarizes the paper.
We use AB magnitude through this work,
as well as a $\Lambda$CDM cosmology with $\Omega_\text{M}=0.3$, 
$\Omega_\Lambda=0.7$ and $H_0=70\text{km s}^{-1}$.

\section{The Quasar Sample} \label{sec:data}

Our input parent quasar sample is based on the V\'eron Catalog of Quasars and AGN, 13th edition
\citep[hereafter the V\'eron Catalog]{veron10}, the SDSS
Data Release (DR) 7 \citep{sch10}, DR12 \citep{paris17} and DR14 \citep{paris17b} quasar catalogs.
SDSS quasar catalogs provide $i$-band absolute magnitudes 
 that is {\it K}-corrected to $z=2$
\citep[$M_i(z=2)$, see][]{richards06} of quasars, 
and the V\'eron Catalog provide $B$ band absolute magnitude ($M_B$).
We convert $M_B$ of quasars from the V\'eron Catalog to $M_i(z=2)$ 
according to the relation in \citet{richards06}.
Quasars which have $M_i(z=2)<-23$  are selected as our parent sample.
We further exclude $z<0.3$ objects to avoid AGNs with extended emission 
which are confusing in PSF subtraction,
and $z>3$ objects because our control sample becomes incomplete at $z>3$
(see \S \ref{sec:maglim} for details).

We use archival broad-band images of the Advanced Camera for Surveys 
Wide Field Camera (ACS/WFC) for quasar companion detection.
The reason to use ACS/WFC images is the small PSF size and large field of view (FOV). 
A small PSF is crucial for detecting companions that are very close to bright quasars,
and a large FOV ensures a valid estimation for the number density of
foreground and background objects.
The ACS/WFC images that contain the selected quasars
are fetched from the Hubble Legacy Archive (HLA)
\footnote{\url{https://hla.stsci.edu/}}.
These images are generated by the HLA using {\texttt{DrizzlePac}} tools \citep{drizzle}, 
which combine raw exposures with the same filter, same camera, and within the same visit.
For each quasar, we choose the deepest image in each band
to form our master image sample. 
In total,  595 quasars are found to appear in 806 images at this step.
We run {\texttt{SExtractor}} \citep{bertin96} on each image to generate a source catalog.

Though image coaddition can enhance the depth of images, 
we do not perform image coaddition because it will introduce
difficulty to the background object number density estimation. 
In most cases, the overlapping area of different images containing the 
same quasar is a small part of the original images. 
The number density of background/foreground objects is estimated
based on the number of objects in the whole image 
(see \S \ref{sec:psfsub} for details). 
Most co-added images do not have enough area to perform a reliable 
background object density estimation.
%{\bf not sure what this means}

We measure the statistics of quasar companions by 
counting all the projected companions and subtracting a background object density.
We thus exclude all quasars that are strongly lensed, 
because the lensed images can be confusing when counting companions.
We further exclude all images that satisfy any of the following:
\begin{enumerate}
\item{Images with NCOMBINE $=1$, where NCOMBINE is the ``NCOMBINE''
	parameter in the header of the {\em HST} image, representing the number of images used for
	cosmic ray rejection when combining raw exposures. Images with NCOMBINE $=1$
	are severely polluted by cosmic rays and are not suitable for companion counting.}
\item{Images where the target quasar is located close (less than $2''$) to
	the edge of the CCD.}
\item{Images that are very crowded. 
%Those images can dominate
	%both projected quasar companion density, background object density
%	and their errors. 
The statistical errors on background object density are high. 
    Images which have more than 5000 objects that were brighter than 25 mag
	in the observed band are excluded in the further analysis.}
\end{enumerate}

\begin{deluxetable}{l|c|c}
\tablecaption{The number of images excluded by the selection criteria.\label{tab:imgsum}}
\tablehead{\colhead{Criteria} & \colhead{Number of Images} & \colhead{Number of Quasars}}

\startdata
Total & 806 & 595\\\hline\hline
Lensed & 6 & 5\\
NCOMBINE $=1$ & 32 & 30 \\
Close the Edge & 25 & 23 \\
Crowded & 63 & 45\\\hline
All Bad Images & 126 & 98\\\hline\hline
Good Images & 687 & 532 \\\hline
\enddata
%\tablenotetext{1}{The number of quasars contained in the corresponding subset of images}
\tablecomments{There are overlaps between different subsets of images / quasars. For example,
one quasar may appear in both bad images and good images. As a result, the total number of quasars 
does not equal to the number of quasars in ``good'' images plus those in ``bad'' images.}
\end{deluxetable}

\begin{figure}
\centering
\includegraphics[trim={0.5cm 0 0 0}, width=3.5in]{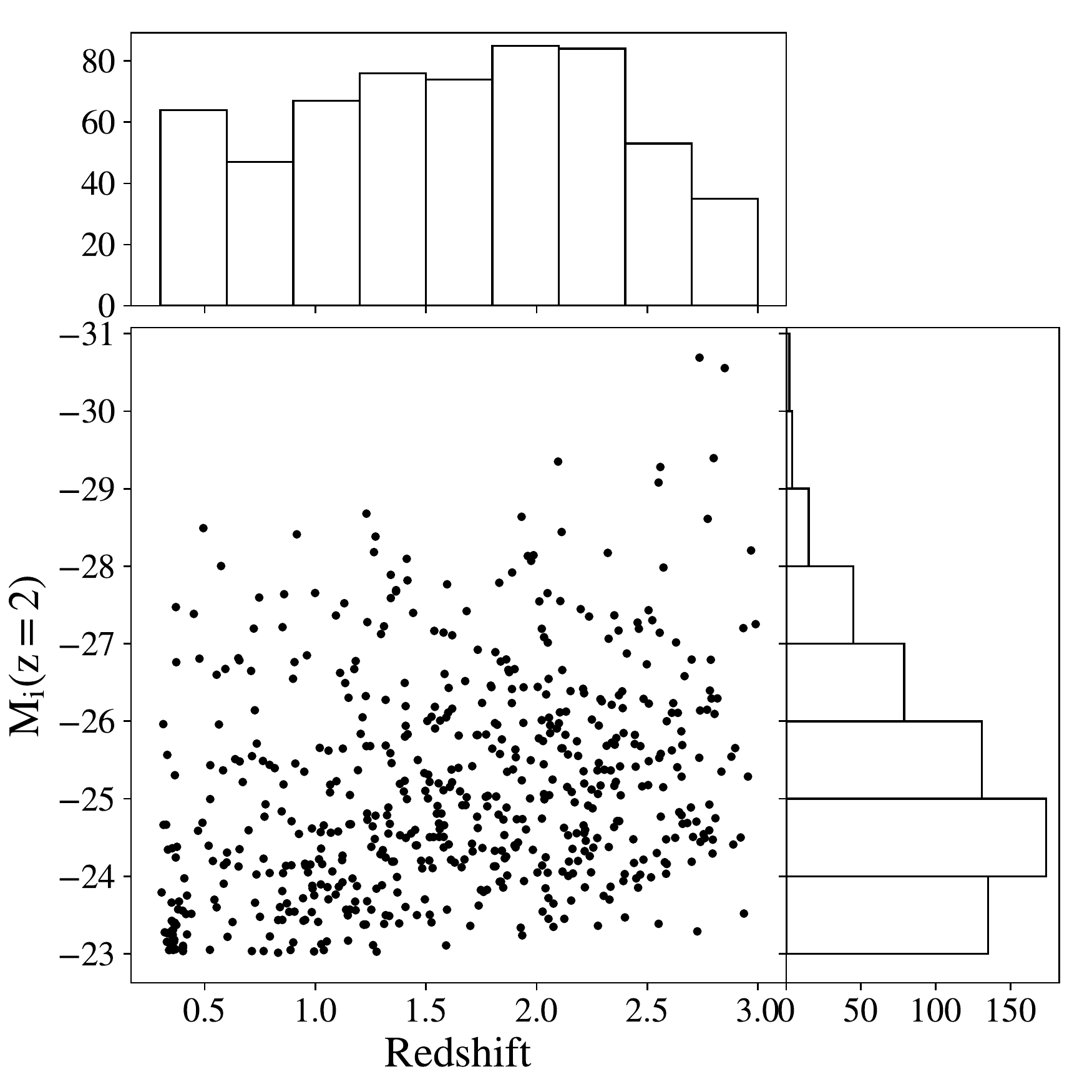}
\caption{The redshift and luminosity distributions of the quasars 
in our master sample. This sample contains 532 quasars which show
up in 687 HST ACS/WFC archival images.}
\label{fig:zmi}
\end{figure}

\begin{figure}
\centering
\includegraphics[trim={0.5cm 0 0 0}, width=3.5in]{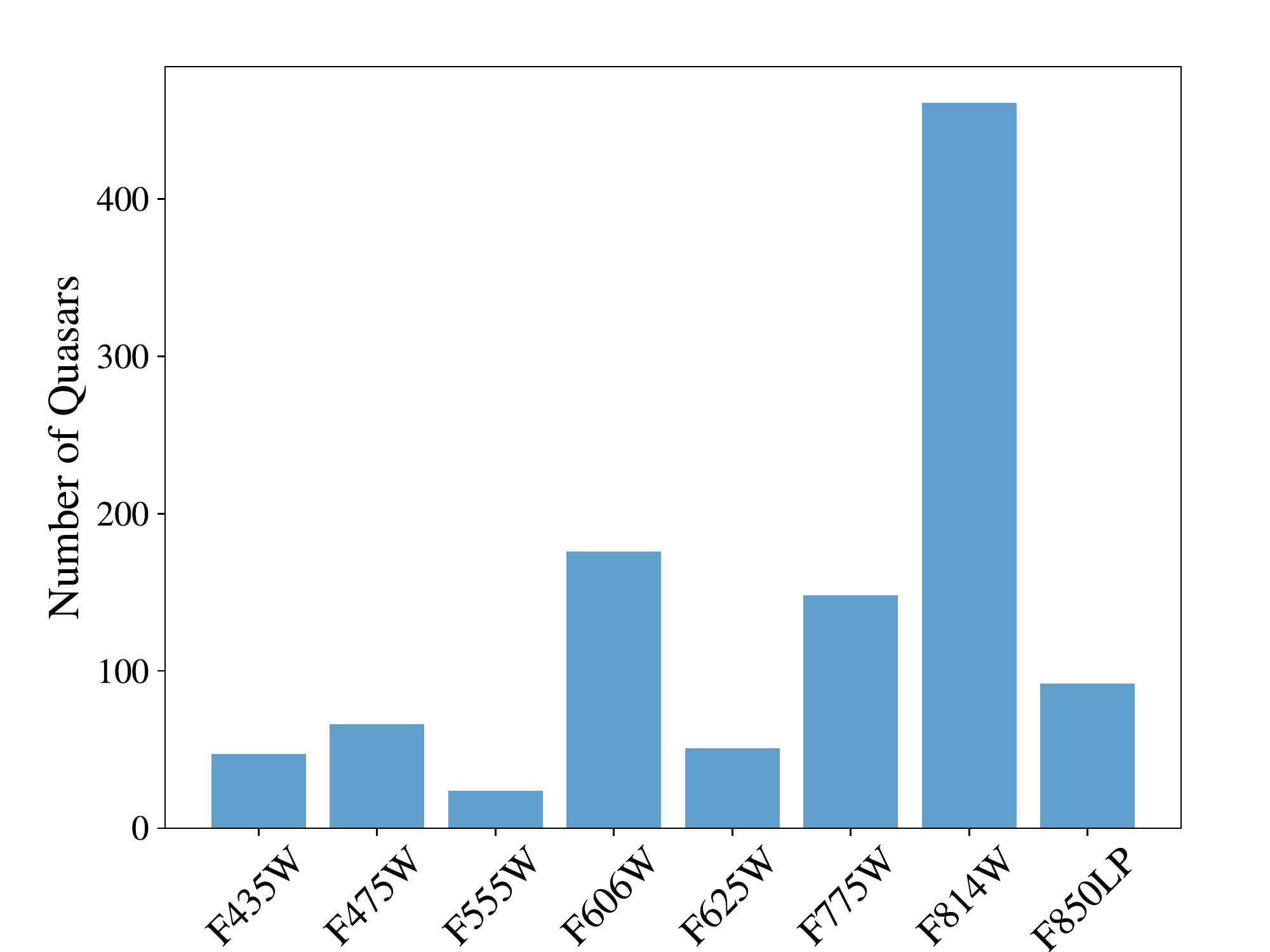}
\caption{Number of quasars observed in each band. 
Note that one quasar can be observed by multiple bands.}
\label{fig:bands}
\end{figure}

We summarize the number of images excluded by each criterion in Table \ref{tab:imgsum}.
All the images that remain after the selection are referred to as ``good'' images in rest of the paper.
The final master sample contains 532 quasars in 687 good images.
Among the 532 quasars, 402 of them were observed in
programs that were not related to AGN studies. 
The fact that most quasars were observed by chance ensures
a small selection effect (see \S \ref{syserr} for further discussion).
In the master sample, one quasar might show up in multiple bands, but there will only be one
image of the quasar given a certain filter.
%{\bf do you double count the companions?}
Figure \ref{fig:zmi} shows the redshift and luminosity distribution of the quasars,
and Figure \ref{fig:bands} shows the number of quasars observed in each band.
Images in the F814W band dominate the sample.

When studying quasar companions of certain magnitude, we will only analyze images 
that are deep enough to detect the faintest companions of interest at a $5\sigma$ level. 
%Correspondingly, one magnitude limit will define a unique quasar sample. 
Previous studies   \citep[e.g.,][]{mat14} showed that
quasar host galaxies have a typical stellar mass range $M_*\gtrsim 10^{10}M_\odot$
and a typical luminosity $M_g\lesssim-21$ in SDSS $g$-band.
We are mainly interested in companions that have stellar masses close to the 
quasar host galaxy, which correspond to major mergers.
We use two sets of magnitude limits:
\begin{enumerate}
\item A ``simple'' cut on absolute magnitude in each band. Companions with 
$M_\text{abs}<-19$ will be analyzed, regardless of in which band the quasar was observed. 
This sample is constructed as a ``pure" observational result that can be directly compared with simulations
without any further assumption.
We convert the absolute magnitude $M_\text{abs}=-19$ to an
apparent magnitude $m(M_\text{abs}=-19)$ 
assuming that the companions have the same redshift as the quasar, 
and require that the image is deep enough to detect an object as faint as $m(M_\text{abs}=-19)$.
We do not perform a {\it K-}correction when converting $M_\text{abs}$ to $m(M_\text{abs})$,
because most quasars were observed in only one band.
If a quasar is observed in multiple bands, 
the deepest image relative to $M_\text{abs}=-19$ is used.
Specifically, the depth of an image relative to $M_\text{abs}=-19$
(regardless of the filter) is quantified by $m_\text{lim}-m(M_\text{abs})$,
where $m_\text{lim}$ stands for the $5\sigma$ depth of the image.
The quasar sample
corresponding to this magnitude cut is referred to as ``sample A." The analysis of this sample will be 
described in \S \ref{sec:psfsub}.
\item A cut based on stellar masses of companions. We convert stellar masses $(M_*)$
to observed magnitudes (denoted by $m(M_*)$) using abundance matching (see \S \ref{sec:maglim} for details).
This sample is constructed to enable physical interpretations of our result.
Companions brighter than $m(10^{9}M_\odot)$ will be analyzed. 
Similar to Sample A,
if a quasar is observed in multiple bands, 
the deepest image relative to $m(10^{9}M_\odot)$ will be used.
The quasar sample
corresponding to this magnitude cut is referred to as ``sample B." The analysis of this sample will be 
described in \S \ref{sec:control}.
\end{enumerate}
The properties of the two quasar samples, together with the master sample, are summarized in Table \ref{tab:sample}.
Unlike in the master sample where a quasar may show up
in multiple images, in sample A and B, 
one quasar only shows up in one image.
%{\bf not clear what this means.}

\begin{deluxetable}{c|l|c}
\tablecaption{Summary of samples used in this study.\label{tab:sample}}
\tablehead{\colhead{Sample} & \colhead{Description\tablenotemark{1}} & \colhead{Number of Quasars}}

\startdata
Master & The master sample & 532\\\hline\hline
A & The image is deep enough to & 230\\
& detect an object of absolute  & \\
& magnitude $M_\text{abs}=-19$ at   & \\
& the quasar's redshift.  & \\
\hline
B & The image is deep enough to & 354\\
& detect an object of apparent  & \\
& magnitude $m=m(10^9M_\odot).\tablenotemark{2}$  & \\
\hline
\enddata
\tablenotetext{1}{All magnitude limits are for point sources at $5\sigma$ level.}
\tablenotetext{2}{The expression $m(M_*)$ refers to the apparent magnitude corresponding to 
stellar mass $M_*$. See \S \ref{sec:maglim} for the detailed definition.}
%\tablecomments{.}
\end{deluxetable}

\section{Detecting Close Companions of Quasars by PSF Subtraction} \label{sec:psfsub}

In this section, we will describe our PSF subtraction and companion detection method.
We will also discuss the statistics of quasar companions with absolute magnitude
$M_\text{abs}<-19$, regardless of the band in which the quasar image was taken.
 Accordingly, we use Sample A throughout this section.

\subsection{Method}
We use close companions of quasars to trace 
the merging history of quasar host galaxies.
If the AGN activity appears in a certain stage of a galaxy merger,
the number of companions around quasars will be different from that of
inactive galaxies.
Specifically, if two merging galaxies are still distinguishable 
when the quasar appears, we will see a pair of galaxies in the field;
if the quasar emerges when the two progenitor galaxies have already 
merged into one galaxy, we will see one single quasar host galaxy rather than a pair.
We will discuss this point in \S \ref{sec:triggermodel} in more details.

%{\bf so I am not sure how to phrase this: was your conclusion in the end that a lack of companion is an indicator of mergers?}
We first perform PSF subtraction to suppress the influence of quasar light on detecting their companions.
The PSF model is generated by \texttt{TinyTim} \citep{tinytim}. \texttt{TinyTim} takes the 
observation time, the position on the CCD chips and the spectrum shape of the source as input parameters.
We use a power-law spectrum with a power-law index of $-2$ ($F_\lambda \propto \lambda^{-2}$)
as the input spectrum to mimic the typical SED of a quasar.
We use a modeled PSF rather than an empirical PSF because 
 most quasars in our sample come from programs in which  the quasars 
are not the primary science targets, and no separate PSF star observations are available.
In addition, the wings of bright quasars can make it difficult to detect projected companions that are 
several arcseconds away from these quasars, thus a large PSF image is preferred.
A modeled PSF can be as large as $20''$, which is difficult  for an empirical PSF.
The size of the PSF models used in this study is $20''\times20''$.

We use \texttt{GALFIT} \citep{galfit} to perform PSF subtraction.
Quasar images are fitted by a PSF component %(mimicking the central engine of the AGN)
plus a S\'ersic profile for the host galaxy.
Examples of PSF subtraction can be found in Appendix \ref{ap1}.
We run \texttt{SExtractor} on the PSF-subtracted images and select all objects 
with a projected distance to the quasar of 
$10\text{ kpc}<d<100 \text{ kpc}$ as projected quasar companions.
We divide the projected distance range $10\text{ kpc}<d<100\text{ kpc}$ into 9
 bins, with the $i$th bin at $10\times i \text{ kpc}<d<10\times (i+1) \text{ kpc}$.
The region where $d<10$ kpc is severely influenced by the PSF-subtraction residuals for many quasars
and is not analyzed.

Companions selected in this way will be inevitably 
contaminated by foreground and background objects.
In the following text, we use ``projected companions''
to represent all the companions detected 
(both physical and unphysical).
To estimate the numbers of physical companions,
a ``background density" representing the background/foreground 
object surface density is calculated for each quasar.
We use the entire image to estimate the background object density.
Specifically, the surface number density of physical companions of the $i$th bin is estimated by
\begin{align}\label{eq:sigma_phys}
\sigma_\text{phys}(d_i, d_{i+1}) = \frac{N(d_i, d_{i+1})}{A(d_i, d_{i+1}) C(d_i, d_{i+1})} - \frac{N_\text{bkg}}{A_\text{bkg}}
\end{align}
where $N(d_i, d_{i+1})$ stands for the number of projected companions with
$10\times i \text{ kpc}<d<10\times(i+1) \text{ kpc}$,
$A(d_i, d_{i+1})$ is the corresponding area, 
and $C(d_i, d_{i+1})$ is the completeness of companion detection
(see \S \ref{sec:completeness} for details).
$N_\text{bkg}$ and $A_\text{bkg}$ describes the numbers of objects
and the area of the whole image.
%We then determine the value of $\sigma_\text{phys}$ for each bin.
The number of physical companions with $10\times i \text{ kpc}<d<10\times j \text{ kpc}$
$(1\le i < j \le 9)$
is calculated by summing up the number of physical companions in the corresponding bins:
\begin{align}
N_\text{phys}(d_i, d_j) = \sum_{i\le k < j}\sigma_\text{phys}(d_k, d_{k+1})A(d_k, d_{k+1})
\end{align}
and the error of $N_\text{phys}(d_i, d_j)$ is estimated assuming a 
Poisson distribution for $N(d_i, d_{i+1})$ and $N_\text{bkg}$ in Eq. \ref{eq:sigma_phys}.
%The number of physical companions around quasars in a certain distance range 
%can be calculated according to the physical companion number density.}

Unless specified, in the rest of the paper, 
``number of companions" refers for the estimated number of physical companions
and ``distance" means projected distance.

%\begin{deluxetable*}{c|cccccccc}
%\tablecaption{Information of the quasar companions.\label{tbl:companions}}
%\tablehead{
%\colhead{Quasar Name} & \colhead{$\text{RA}_\text{Q}$\tablenotemark{a}} & %\colhead{$\text{DEC}_\text{Q}$} 
%& \colhead{Redshift} & \colhead{Filter} & \colhead{$\text{RA}_\text{C}$\tablenotemark{b}} & 
%\colhead{$\text{DEC}_\text{C}$} &  \colhead{$m_\text{companion}$} & \colhead{Distance} \\
% \colhead{} & \colhead{(deg)} & \colhead{(deg)} & \colhead{} & \colhead{} & \colhead{(deg)} 
% & \colhead{(deg)} & \colhead{} & \colhead{($"$)}
%}
%\startdata
%SDSS J1001+0200 & 10:01:51.09 & 02:00:31.23 & 0.967 & F814W & 10:01:51.42 & 02:00:36.21 & 24.1 & 7.01 \\
%                &             &             &       &       & 10:01:51.12 & 02:00:23.79 & 23.6 & 7.45 \\
%                &             &             &       &       & 10:01:51.49 & 02:00:26.05 & 23.9 & 7.81 \\
%                &             &             &       &       & 10:01:51.78 & 02:00:30.74 & 20.5 & 10.3 \\
%                &             &             &       &       & 10:01:50.41 & 02:00:34.80 & 24.3 & 10.8 \\
%\enddata

%\tablenotetext{a}{The coordinates of quasars.}
%\tablenotetext{b}{The coordinates of quasar companions.}
%\tablecomments{The full
%table is available at (someaddress)}
%\end{deluxetable*}

\subsection{Companion Detection Completeness} \label{sec:completeness}

Some companions may be missed or misidentified as a result of 
imperfect PSF subtraction % by simulating fake companions.
because it can be difficult to distinguish companions from the PSF-subtraction residuals.
%{\bf I don't think you ever discussed false-positives as a result of PSF residuals.}
The probability of a companion to be detected 
is mainly influenced by the flux contrast between the companion and
the PSF-subtraction residual.
We find four factors that have a major influence on the completeness:
(1) the flux of the quasar, (2) the accuracy of the PSF models,
(3) the angular distance from the quasar to the companion,
and (4) the flux of the companion.
Factors (1) and (2) vary from quasar to quasar, 
while factors (3) and (4) are determined by the companion itself.
Accordingly, we add simulated companions 
with different flux and distance for each quasar image
 to estimate the fraction of missed companions. 
The simulated companions are generated to be point sources. 
 For each quasar image, we simulate three sets of companions, with 
absolute magnitude $M_\text{abs}=-19, -20, -21$
in the corresponding band,
assuming the companions to have the same redshift as the quasar.
The completeness is estimated as a function of projected distance to the quasar.
%We divide the distance range $10\text{kpc}<d<100\text{kpc}$ into 9
% bins, with a bin width of $\Delta d=10$kpc, 
We generate 10 simulated companions at
randomized positions in each distance bin.
The simulated images are analyzed by the companion detection process described above,
according to which we calculate the completeness of the companion detection for each quasar.
A simulated companion is regarded as ``detected"
if an object is detected within $0\farcs2$ from the position of the simulated companion.

In the simulation, 
we do not consider false positives resulting from the PSF-subtraction residual
because the probability for a PSF-subtraction residual to appear right at 
the position of a simulated companion is negligible. 
The simulation ensures that most companions of interest can be detected.
To exclude false positives in the real images,
we visually inspect all the images and remove suspicious detections 
that are likely PSF-subtraction residuals. 

%{\bf you should add a few sentences describing your choice of the lower distant limit, i.e., why you do not consider things within 10 kpc or so.}

\begin{figure}
\centering
\includegraphics[width=3.3in]{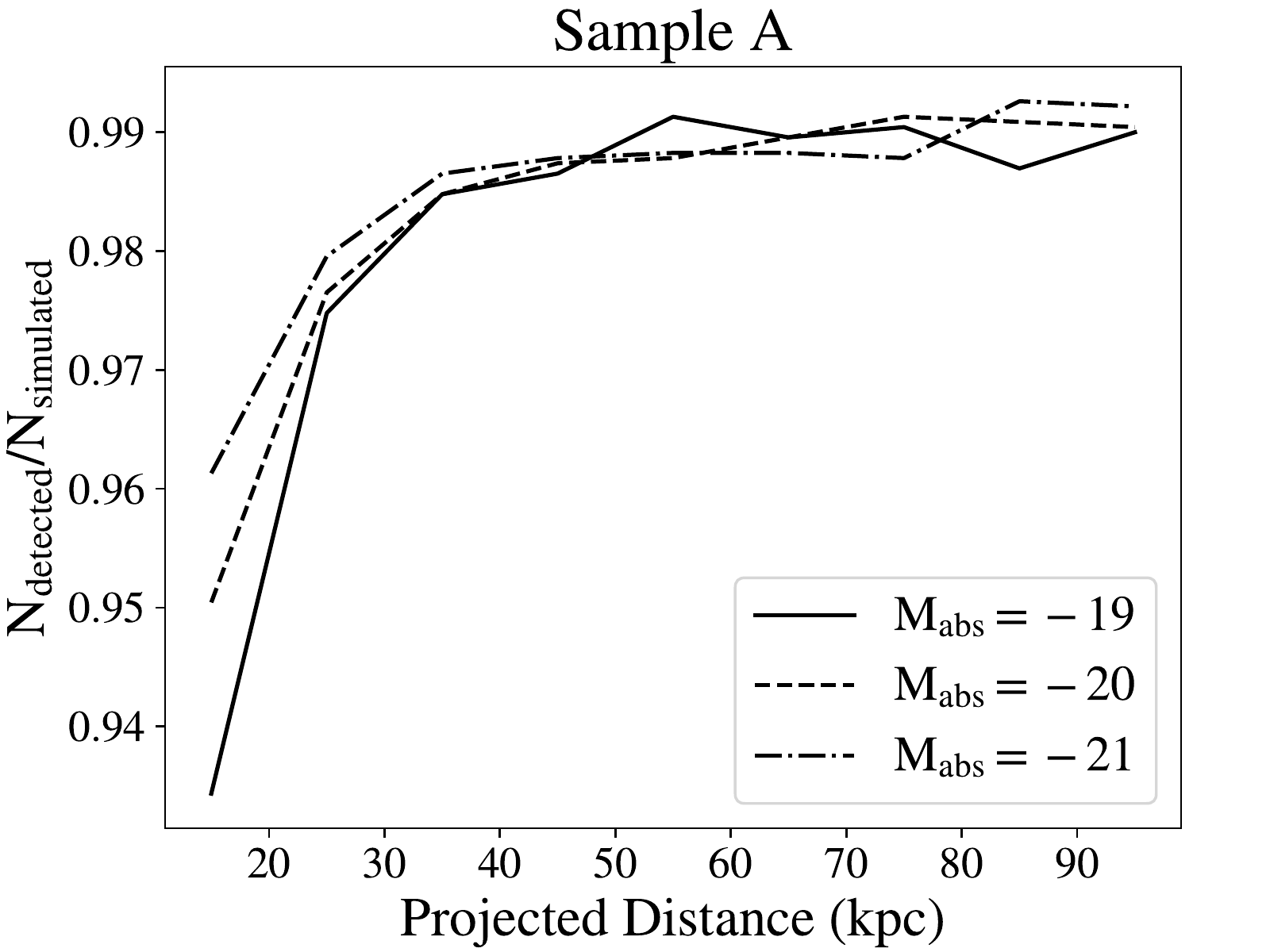}
\caption{The completeness of companion detection estimated by simulated companions.
The simulated objects are generated as point sources.
We estimated that the fraction of missed companions is less than 10\%
for all the companions of interest.}
\label{fig:completeness}
\end{figure}

Figure \ref{fig:completeness} shows the average completeness 
 as a function of companion flux
and the distance to the quasar from our simulation.
The completeness is larger than 90\% even for the faintest companion
in the smallest distance bin $(10\text{ kpc}<d<20\text{ kpc})$.
 For companions that are more than 50 kpc away from 
 the quasar (which corresponds to $\sim 6''$ at $z=1.6$), the completeness
 does not change with the distance, which indicates that
 the influence of the quasar light becomes negligible at larger distances.

We test the potential influence of using point sources as simulated companions.
We run another simulation where the shape of the companions are exponential disks with
an effective radius of $r_e=2\text{ kpc}$, assuming that the companion has a same redshift as the quasar.
The difference between the completeness given by the two simulations is less than 1\%.
%{\bf I have a previous comment about false-positives, i.e., do you ever detect companions where there is none as a result of PSF residual. I don't think you have addressed that.}

\subsection{Quasar Companion Fraction\footnote{The information of all the quasars and quasar companions
is available at (\url{https://github.com/yuemh/qso_companion})}} \label{sec:primary_result}

\begin{figure}
\centering
\includegraphics[width=3.3in]{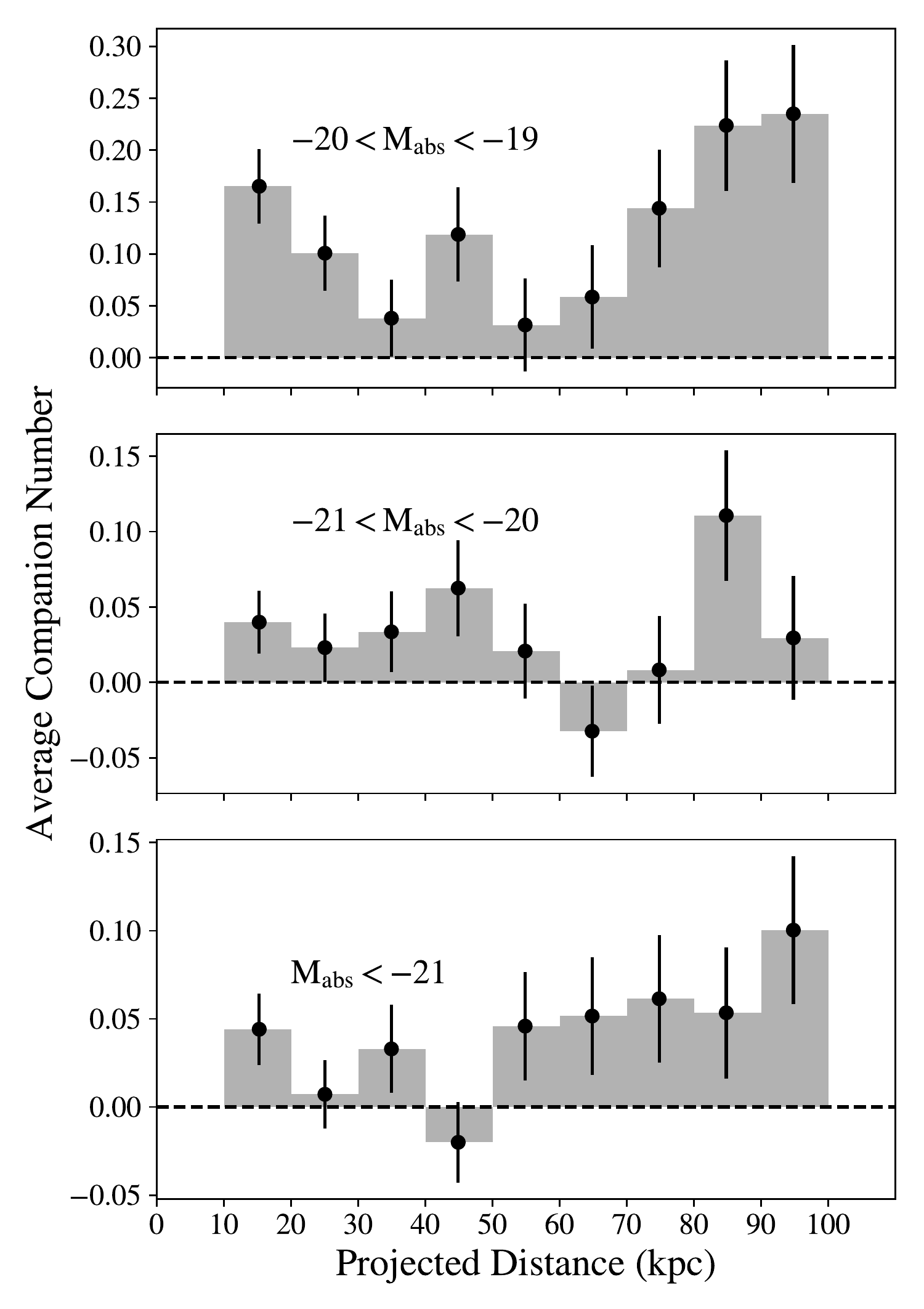}
\caption{The average companion numbers of sample A quasars.
Each data point corresponds to a distance bin of $\Delta d=10$ kpc.
All the error bars represent $1\sigma$ error assuming a Poisson distribution for
the number of companions (same for all the other figures).
}
\label{fig:primary}
\end{figure}

Here we examine the statistics of quasar companions 
with $-20<M_\text{abs}<-19$, $-21<M_\text{abs}<-20$
and $M_\text{abs}<-21$, regardless of the band in which
the quasar was observed.
We do not perform $K-$ corrections on the magnitudes.
Figure \ref{fig:primary} shows the average number of companions around quasars in Sample A.
The number is negative in some distance ranges due to the subtraction
of background object number density.
%The excess of number density of $-21<M_\text{abs}<-19$ companions
% mainly appears at $d<30$kpc and $d\sim 80$kpc,
% where $d$ is the projected distance to the quasar.
We define companions with a projected distance of 
$10 \text{ kpc}<d<30 \text{ kpc}$ to quasars as ``close companions" and calculate
the average number of close companions $(\overline{N}_\text{comp})$ around quasars (galaxies).
 The distance range is chosen because companions at 
$10 \text{ kpc}<d<30 \text{ kpc}$ are likely involved in a merging event, 
while companions at larger distances
are not necessarily associated with mergers.
A similar distance range has been adopted in previous studies on the galaxy merging rates
\citep[e.g.,][]{lambas03, ellison08, DR09, man12}.
% The total number of projected close companions (both physical and unphysical) 
% around the 230 quasars is 639.
 The average numbers of close physical companions are 
 $0.26\pm0.05$, $0.06 \pm 0.03$, and $0.05 \pm 0.03$  for $-20<M_\text{abs}<-19$,
 $-21<M_\text{abs}<-20$ and $M_\text{abs}<-21$ companions,
 and $0.38 \pm 0.07$ for all physical companions which have $M_\text{abs}<-19$.

One issue of this analysis is that the ``absolute magnitude''
we use here is difficult to be translated to physical properties
of companions, given that we do not perform $K$-corrections on these magnitudes.
{\it K}-corrections require knowledge of the companion SED, 
while we usually have only one-band measurement. This issue will be addressed in \S \ref{sec:control},
by introducing a magnitude cut which is associated with galaxy stellar masses.

\subsubsection{On-going Merging Systems} \label{subsec:candidates}
In addition to statistical studies of large samples,
detailed modeling and observations of individual cases are
also crucial to the understanding of the quasar triggering mechanism.
Here we report some on-going mergers
with quasar activity. Follow-up observations on these objects,
such as host galaxy morphology, gas kinetics and AGN obscuration
can be compared directly with simulations. 
We visually inspect images of quasars in the master sample,
and select objects that show disrupted features like tidal tails and asymmetric host galaxies.
We find 22 quasars with features of recent mergers.
The images of these objects can be found in Appendix \ref{ap1}.
Note that these objects are included in both sample A and B.

\section{Comparing the Companion Fraction of Quasars with a Galaxy Control Sample} \label{sec:control}

To compare the companion distribution of quasar host galaxies and normal galaxies,
we use the 3D-HST galaxy catalog \citep{brammer12,skelton14,mom16} 
to construct a control sample of galaxies.
The 3D-HST is an {\em HST} Treasury program to provide ACS and WFC3 images and grism spectroscopy
over five fields: COSMOS, GOODS-north, GOODS-south, AEGIS, and UDS
with a combined usable area of $\sim 0.25$ deg$^2$.
In this work, we use the 3D-HST photometry catalog v4.1.5, which is the latest version.
It provides ACS and WFC3/IR broad-band fluxes of galaxies,
including F160W, F140W and F125W for WFC3/IR images,
as well as F814W and F606W for ACS images.
The catalog also contains stellar mass and photometric redshifts of galaxies, 
derived with methods described in \citet{skelton14}.

Quasar host galaxies are believed to be massive 
\citep[$ M_* \gtrsim 10^{10} M_\odot$, e.g.][]{dunlop03, mat14}.
We thus construct a ``massive galaxy sample" by selecting all the galaxies 
that have $M_*>10^{10}M_\odot$ and photometry flag {\texttt{USE\_PHOT}}$=1$ 
(which means good photometry) in the 3D-HST photometry catalog.
The control sample galaxies are randomly drawn from the massive galaxy sample
and share the same redshift distribution as our quasars.
This is done by evenly dividing the redshift range ($0.3<z<3$) into 9 bins
and randomly selecting galaxies in each redshift bin, so that the fraction of galaxies
in a specific redshift bin among all the control sample galaxies equals to the fraction of
quasars in that redshift bin among all the quasars.

All projected companions of the control sample galaxies
with a projected distance $10$ kpc $<d<100$ kpc
are selected in the same way as quasars. 
Since the 3D-HST contains five fields with different depths,
the ``background object density" of a control sample galaxy
is approximated by the object surface number density of 
the 3D-HST field where the galaxy is located.
We use F814W magnitudes to make magnitude cuts when counting companions,
since F814W images dominate our quasar sample (Figure \ref{fig:bands}).

\subsection{Making Absolute Magnitude Cuts Based on Stellar Mass} \label{sec:maglim}
In \S \ref{sec:primary_result}, 
we show the number of companions around quasars as a function of companion flux, 
where the flux of companions are described by absolute magnitudes without {\it{K}}-corrections.
This result is not suitable for
studying the redshift evolution of quasar companions or
comparing quasars with control sample galaxies,
since different bands are used in the analysis of the quasars (all the 8 broad bands of 
ACS/WFC) and the galaxies (F814W only).
It is difficult to perform {\it{K}}-corrections in this work,
because most of the quasars have only been observed in one band.
Therefore, we use an abundance matching technique to convert stellar mass cuts
to magnitude cuts in each band, which allows us to set magnitude cuts
consistently at any redshift in all the eight bands.
In short, for a given stellar mass $M_*$,
we find a magnitude (denoted by $m(M_*)$) such that that 
the number of galaxies that are more massive than $M_*$
(denoted by $N(M_*)$) is the same as the number of galaxies that are brighter than $m(M_*)$.
We describe the detailed procedures below.

The analysis is based on the photometric catalog of the UDS field in the 3D-HST project. 
The UDS photometric catalog is used because it contains the Johnson $B,V,R$ and SDSS $i',z'$ photometry,
which can be converted to {\em HST} broad-band magnitudes and be compared with
the quasars. The conversion is done as follows.
The SDSS $i',z'$ magnitudes are converted to Johnson $I$ magnitude according to \citet{jordi06},
then the $BVRI$ magnitudes are converted to ACS/WFC broad band magnitudes
according to \citet{sirianni05}. The conversion from $BVRI$ to F850LP was not provided in \citet{sirianni05},
so we used SDSS $z'$ as an approximation of the F850LP magnitudes. We estimate the difference between
SDSS $z'$ and ACS/WFC F850LP magnitudes using the Exposure Time Calculator for {\em HST} ACS/WFC
using typical galaxy templates. The difference is smaller than 0.02 magnitude.
We select galaxies in the UDS photometric catalog with {\texttt{USE\_PHOT}}$=1$ to construct the galaxy sample for the abundance 
matching technique.
Note that this galaxy sample is different from the control sample; 
we only use the UDS field in the five 3D-HST fields and do not set any 
stellar mass cut on this sample.

For a quasar at redshift $z_q$, the magnitude cut (in an arbitrary band) 
corresponding to a stellar mass $M_*$ is estimated as follows.
First, all objects with photometric redshift ($z_\text{phot}$) 
that satisfy $z_q-0.1<z_\text{phot}<z_q+0.1$ in the UDS catalog are selected, 
constructing a ``magnitude-cut-setting galaxy sample''.
Typically the redshift-matched galaxy sample contains $1000\sim 3000$ galaxies.
%{\bf how many objects are in the sample?}
The magnitudes of the objects in the magnitude-cut-setting galaxy sample
are corrected to $z_q$ according to the photometric redshifts,
i.e., given a galaxy with a photometric redshift $z_\text{phot}$ 
and an apparent magnitude $m_\text{raw}$, we calculate the corrected magnitude by
\begin{align} \label{eq:zphot}
m=m_\text{raw} + 5\log [D_L(z_q)/D_L(z_\text{phot})]
\end{align}
where $D_L(z)$ is the luminosity distance.
We then count the number of objects with stellar masses larger than
the given value $M_*$ (referred to as $N(M_*)$) in the ``magnitude-cut-setting galaxy sample".
The value of $N(M_*)$ varies from several hundred (for $M_*=10^9M_\odot$)
to about twenty (for $M_*=10^{11}M_\odot$).
Finally, we find the magnitude limit $m(M_*)$ 
so that there are $N(M_*)$ objects that are brighter than $m(M_*)$.

\begin{figure}
\centering
\includegraphics[width=3in]{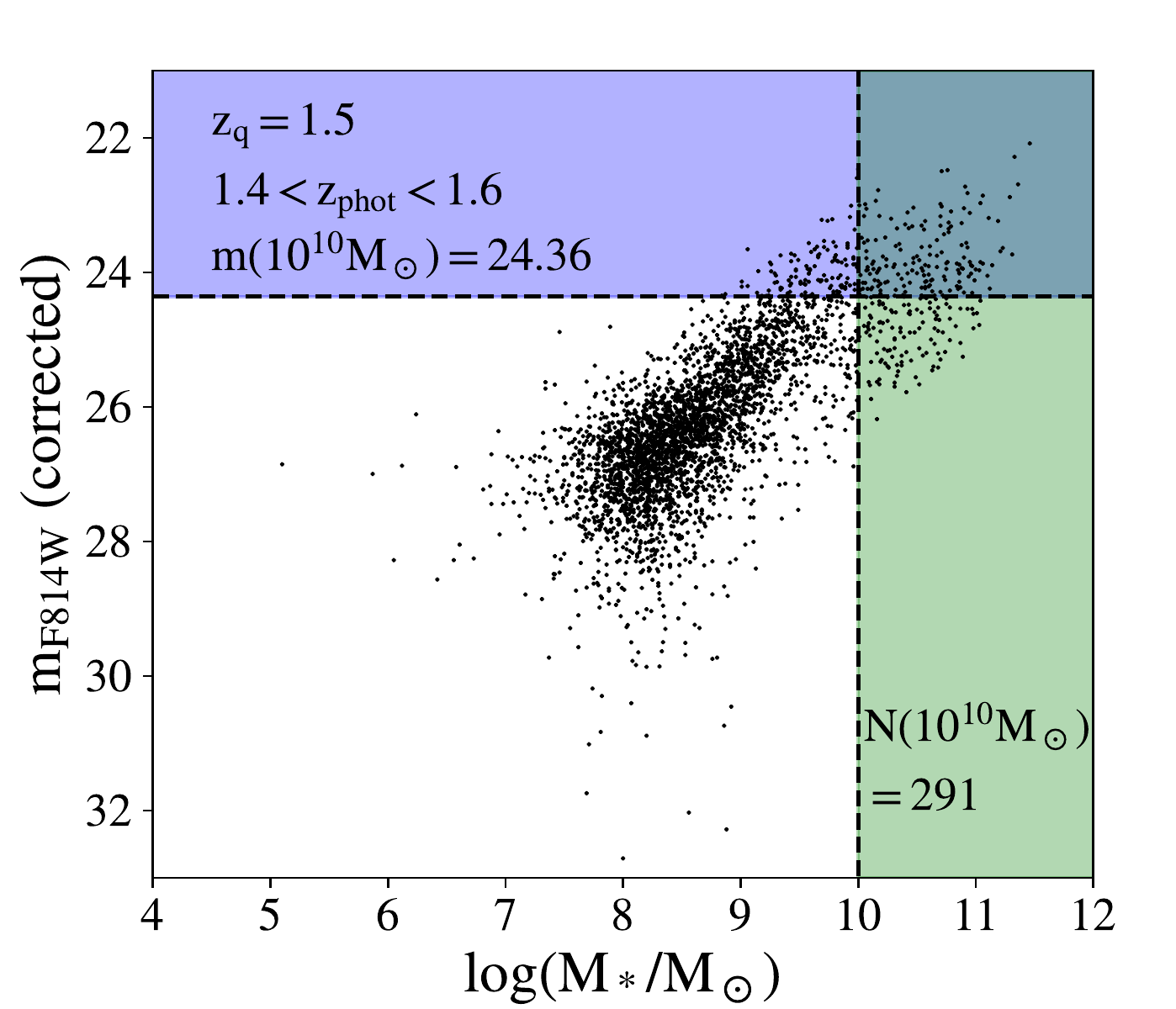}
\caption{An example of converting stellar mass limits into apparent magnitude limits. 
In this example, we convert a stellar mass of $10^{10}M_\odot$ to F814W magnitude for $z\sim1.5$ objects.
We first select objects in the UDS catalog with photometric redshift $1.4<z_\text{phot}<1.6$,
and correct their F814W magnitude according to Eq. \ref{eq:zphot}.
There are 291 objects with $M_*>10^{10}M_\odot$ (objects in the green shade), 
which is also the number of objects with corrected
F184W magnitude brighter than $m(10^{10}M_\odot)=24.36$ (objects in the blue shade).
}
\label{fig:matching}
\end{figure}

As an example, Figure \ref{fig:matching} illustrates the process
of estimating $m(10^{10}M_\odot)$ in F814W band at $z=1.5$.
The number of objects in the green and the blue shade is equal.
By definition, we can
estimate the number of quasar companions with stellar masses larger than
$M_*$ by by counting objects that are brighter than $m(M_*)$.

%{\bf not sure what's the purpose of the following sentence.}
%By definition, we can estimate the number of quasar companions with
%stellar masses larger than $M_*$ by counting objects that are brighter than
%$m(M_*)$.
%{\bf how many objects are usually used in the matching? You should spend a few sentences discussing the potential systematics with this matching approach.}

\begin{figure}
\centering
\includegraphics[width=3.3in]{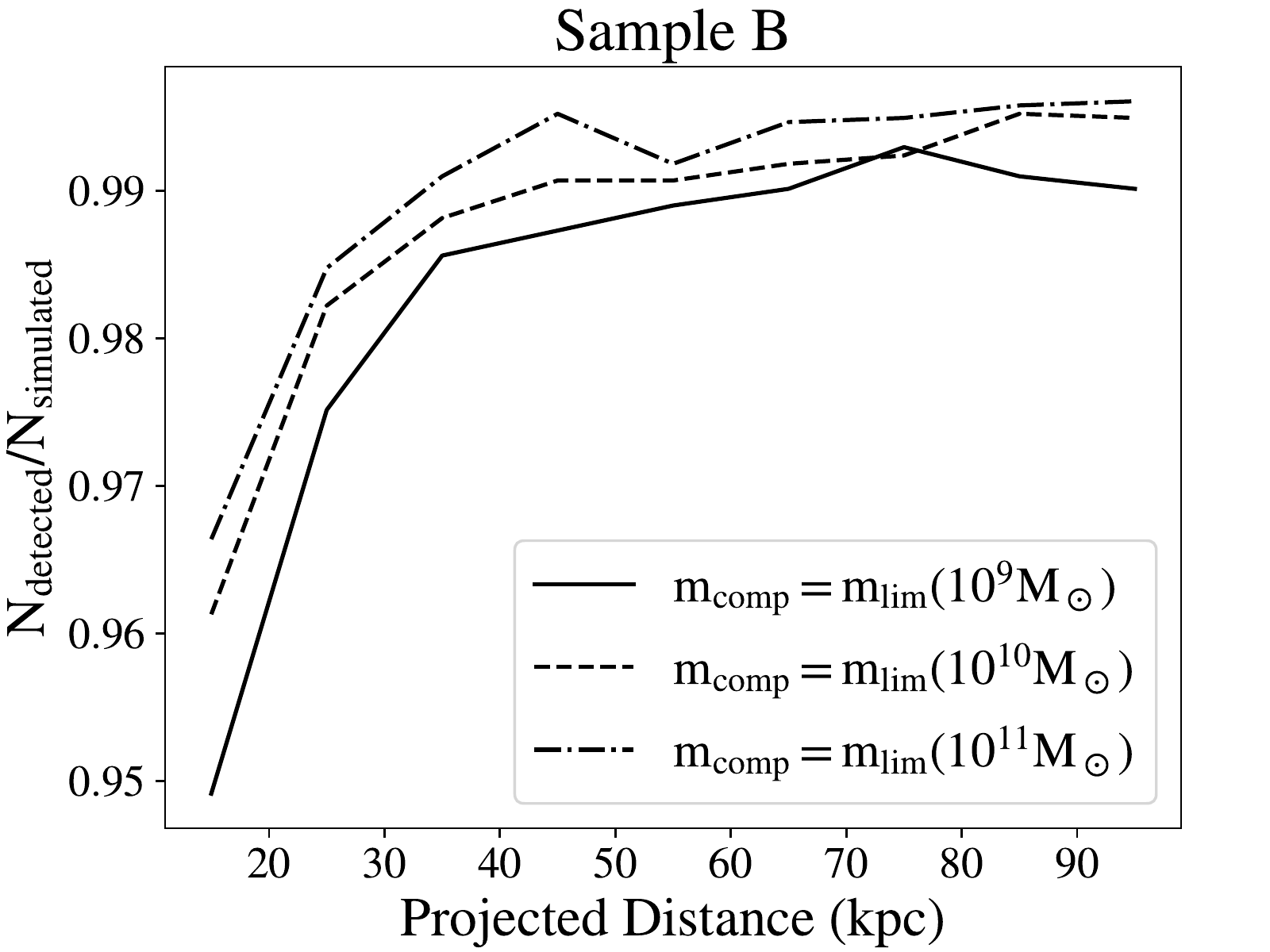}
\caption{The detected fraction of quasar companions which have a magnitude of
$m(10^9M_\odot)$, $m(10^{10}M_\odot)$ and $m(10^{11}M_\odot)$, 
estimated by simulated images.
The detected fraction is higher than 95\% for all the companions of interest.}
\label{fig:completeness2}
\end{figure}

%\textcolor{red}{
%We thus can establish a homogeneous set of magnitude cuts 
%in all the eight bands at any redshift.}
For each quasar image, we set up three magnitude limits,
$m(10^9M_\odot)$, $m(10^{10}M_\odot)$
 and $m(10^{11}M_\odot)$.
When comparing quasars with the galaxy control samples,
we will discuss companions with $m(10^9M_\odot)>m>m(10^{10}M_\odot)$ (faint),
$m(10^{10}M_\odot)>m>m(10^{11}M_\odot)$ (intermediate) and 
$m<m(10^{11}M_\odot)$ (bright). 
Correspondingly, we use Sample B throughout \S \ref{sec:control}.
Since most quasar host galaxies have stellar mass $M_*>10^{10}M_\odot$,
``intermediate'' companions are mainly associated with 
major mergers (mass ratio is close to 1),
while ``faint'' companions are mainly
related to minor mergers (mass ratio is larger than 3).
The case of ``bright" companions is more complicated;
these systems might be associated with minor mergers 
where the quasar host galaxy is the less massive progenitor,
or major mergers if the quasar host galaxy is as massive as $10^{11}M_\odot$.
Studies on quasar host galaxies \citep[e.g.,][]{mat14, yue18}
show that only a small fraction of quasar host galaxies have stellar masses
larger than $10^{11}M_\odot$, thus most systems with bright companions
should be related to minor mergers.
We estimate the completeness of companion detection for companions with
$m(10^9M_\odot)$, $m(10^{10}M_\odot)$
 and $m(10^{11}M_\odot)$, using the same method as described in
\S \ref{sec:completeness}.
The result is presented in Figure \ref{fig:completeness2},
which shows that the completeness of companion detection is higher than
95\% in the simulated images.

Similar magnitude limits are calculated for galaxies in the control sample
using the F814W magnitude.
At $z=3$, the faintest companions that we consider in this study have magnitude
 $m_\text{F814W}(10^9M_\odot)=26.3$.
At this magnitude, about 2\% objects in the 3D-HST catalog have signal-to-noise ratios smaller than 3,
and the completeness of companion counting will drop toward fainter magnitudes
(and thus higher redshift).
This is the reason why our sample only contains $z<3$ objects.

\subsection{Comparison Between Companions around Quasars and Galaxies} \label{sec:results}

\begin{figure}
\centering
\includegraphics[trim={1cm 0 0 0},width=3.3in]{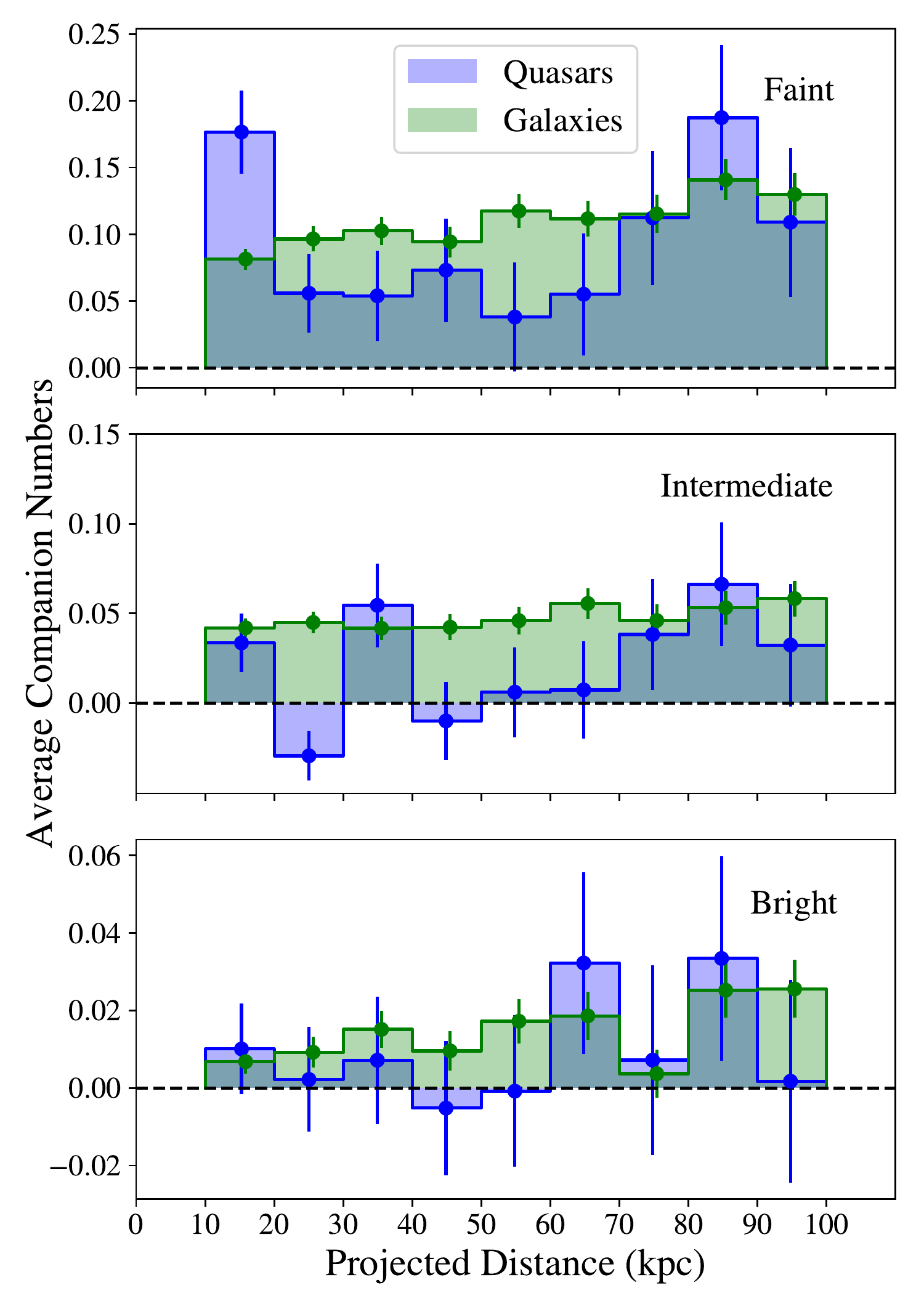}
\caption{The average number of companions of quasars and the control sample galaxies.
Each data point corresponds to a distance bin of $\Delta d=10$kpc.
Small x-axis offsets are added to the error bars to make them distinguishable.
}

\label{fig:density2}
\end{figure}

We calculate the number of faint, intermediate and bright companions of quasars
in Sample B.
Figure \ref{fig:density2} shows the average number of physical companions,
both for quasars and the control sample galaxies.
%, after subtracting the foreground/background contaminants.
 Quasars have fewer intermediate companions at $d<60$ kpc.  
 At larger distances, the number of companions around quasars and galaxies
 are roughly the same.
 No significant difference can be seen for faint and bright companions
 between quasars and galaxies.

%No significant difference can be seen for ``brightest'' companions
% around quasars and galaxies.
% The companion number density profiles of quasars and galaxies 
% are roughly consistent for both ``faint" and ``bright" companions. 
% Among all the distance bins, the largest difference is $\sim 3\sigma$.
% However, there are still some features that worth noticing. First,
% there is a deficit of companions around quasars at $d<30$kpc.
% The difference is of $3\sigma$ significance if we consider all the companions at $10<d<30$kpc.
% We will raise a possible explanation to this deficit in Section \ref{sec:discussion}.

\begin{figure*}
\centering
\includegraphics[trim={1cm 0 0 0},width=7.3in]{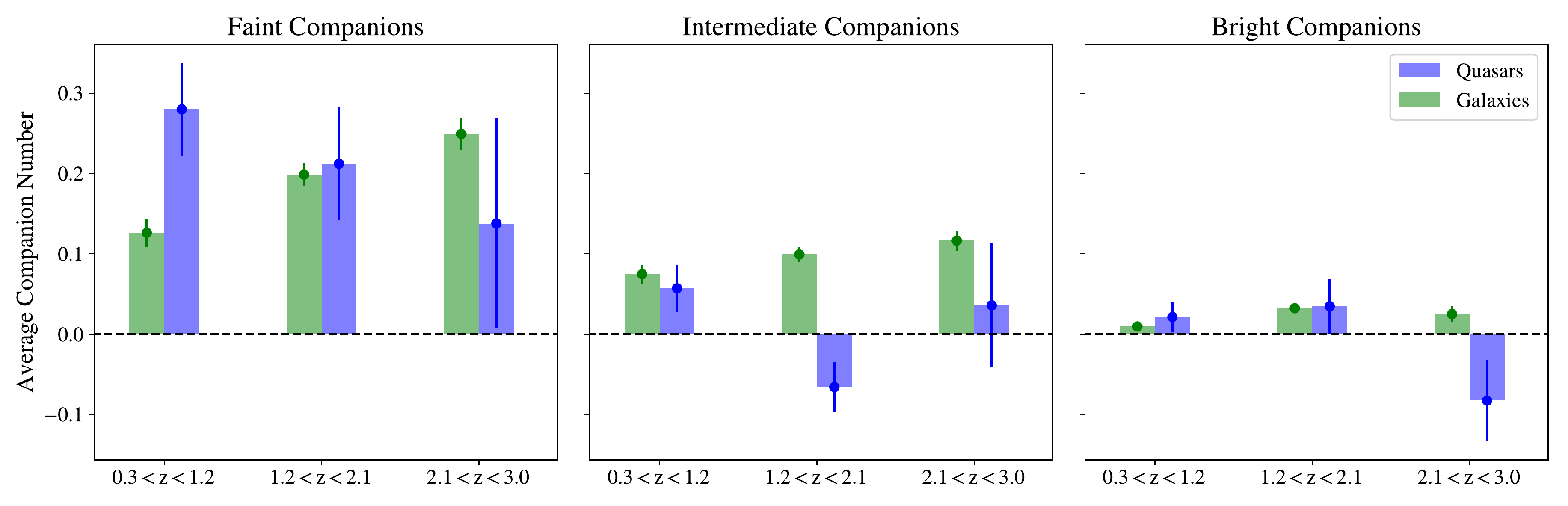}
\caption{The redshift evolution of average number of companions around quasars.
%The three redshift bins correspond to $0.3<z<1.2, 1.2<z<2.1,
%2.1<z<3.0$, respectively.
The average number of companions of galaxies in the control sample is also included.
}
\label{fig:pf_z}
\end{figure*}

\begin{figure}
\centering
\includegraphics[trim={1cm 0 0 0},width=3.3in]{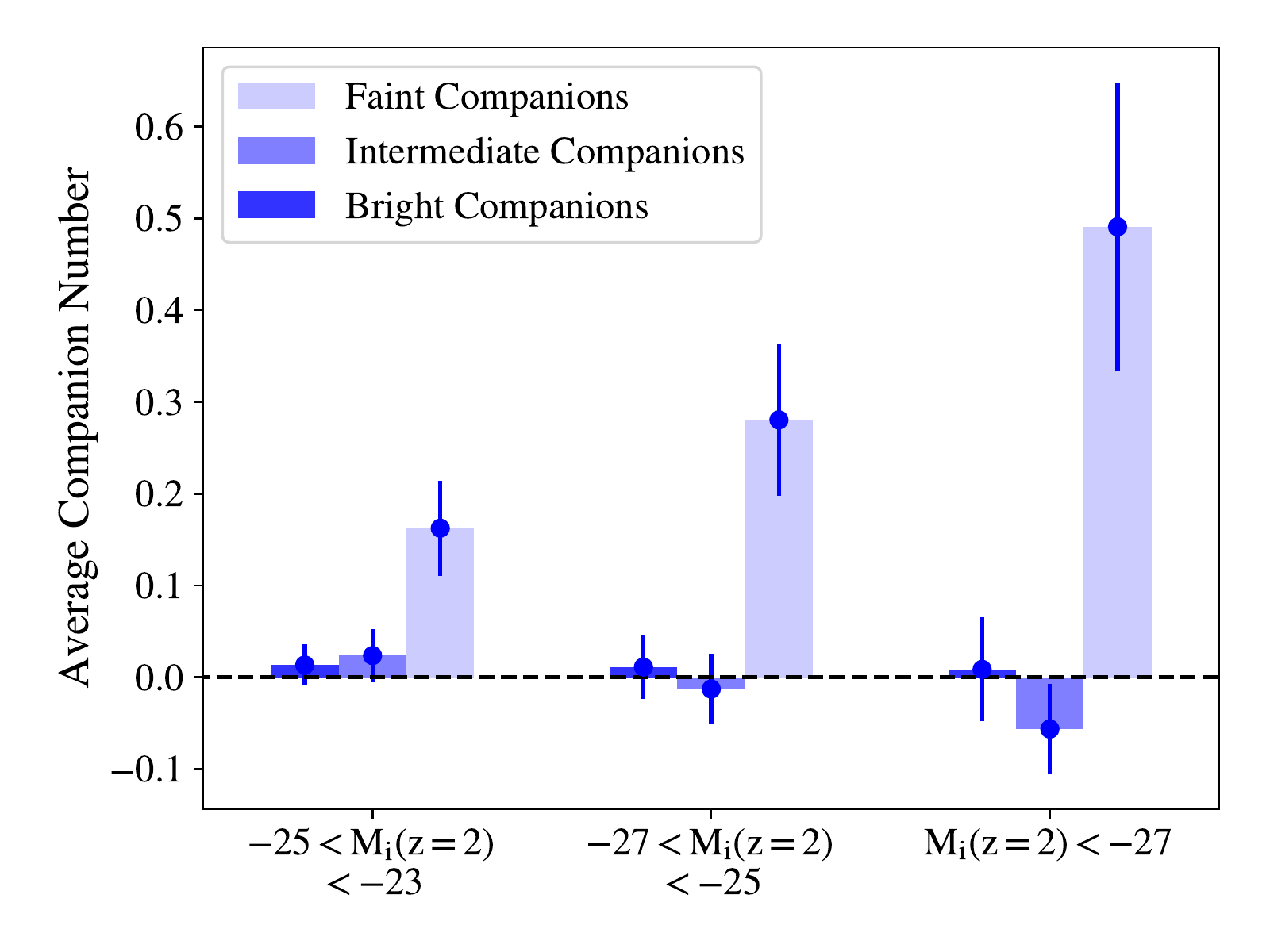}
\caption{The evolution of the average number of companions around quasars with quasar luminosity.
%The three luminosity bins are $-25<M_i(z=2)<-23, -27<M_i(z=2)<-25$,
%and $M_i(z=2)<-27$, respectively.
}
\label{fig:pf_m}
\end{figure}

For Sample B quasars, the average numbers of ``close companions'' (companions with a project distance to quasars of
$10 \text{ kpc}<d<30\text{ kpc}$) are

\begin{align}\label{eq:pfquasar}
&\overline{N}_\text{comp,Q}(\text{faint})=0.233 \pm 0.043 \nonumber \\
&\overline{N}_\text{comp,Q}(\text{intermediate})=0.004 \pm 0.021 \\
&\overline{N}_\text{comp,Q}(\text{bright})=0.012 \pm 0.018  \nonumber
 \end{align}
%\end{multline}

and the comparison galaxy sample has

\begin{align}\label{eq:pfgalaxy}
&\overline{N}_\text{comp,G}(\text{faint})=0.178 \pm 0.012  \nonumber \\
&\overline{N}_\text{comp,G}(\text{intermediate})=0.087 \pm 0.008 \\
&\overline{N}_\text{comp,G}(\text{bright})=0.016 \pm 0.005  \nonumber
 \end{align}
 
Quasars show a $3.7\sigma$ deficit of close intermediate companions, and have 
 numbers of faint and bright companions similar to those of normal galaxies.

Figure \ref{fig:pf_z} shows the average number of close companions of quasars
and control sample galaxies as a function of redshift. 
The average number of companions around control sample
galaxies increases toward high redshift, both for faint and bright companions. 
This result is expected since the universe is more crowded at high redshift.
Meanwhile, the average number of companions around quasars does not significantly evolve
with redshift.

Figure \ref{fig:pf_m} shows the average number of close companions
as a function of quasar luminosity. 
More luminous quasars have more faint companions, while 
the number of intermediate companions decreases with quasar luminosity.
Given the statistical error, both trends are not significant.
The number of bright companions does not show a luminosity dependence.

We also examine the relationship between average companion number of quasars and other quasar properties, including:
\begin{enumerate}
    \item{
    Broad absorption line (BAL) features. The SDSS quasar catalogs contain BAL flags, 
    while the V\'eron catalog does not have such information, 
    thus our BAL and non-BAL quasar sample only contain SDSS quasars.
    We find that BAL quasars and non-BAL quasars have consistent numbers of companions 
    (both faint and intermediate) at a $1\sigma$ level.}
    
    \item{Radio loudness. We match our quasar catalog with the
    {\it Faint Images of the Radio Sky at Twenty-Centimeters}
    \citep[{\it FIRST}; e.g.,][]{first, first2} survey catalog.
    We divide the quasars that were covered by the {\it FIRST} survey into two subsamples,
    namely ``radio loud (RL) quasars'' and ``radio quiet (RQ) quasars'', 
    according to whether they were detected by the {\it FIRST} survey.
    We find that RL quasars tend to have more ``faint" companions than RQ quasars.
    The average number of ``faint" companions is $0.523 \pm 0.144$ around RL quasars
    and $0.153 \pm 0.047$ around RQ quasars, which is a $2.4\sigma$ difference.
    The two samples have similar numbers of intermediate companions (consistent at a $1\sigma$ level).
    }
    
    \item{Infrared (IR) brightness. 
    We match our quasar catalog with the {\it Wide-field Infrared Survey Explorer}
    \citep[{\it WISE}; e.g.,][]{wise} ALLWISE source catalog.
    Since {\it WISE} W3 and W4 are usually not deep enough for faint quasars,
    we use the {\it WISE} W1 and W2 fluxes to estimate the rest-frame $2\mu \text{m}$ magnitude
    ($M_{2\mu}$) assuming that the near infrared SED of quasars can be represented by a power law.
    We then evenly divide the quasars into two subsamples 
    according to their $M_{2\mu}-M_i(z=2)$ color, 
    referred to as ``IR-bright quasars'' and ``IR-faint quasars'', respectively.
    One potential problem is that the near-infrared SED of some quasars cannot be well-fitted 
    by a single power law \citep[e.g.,][]{glickman06, HC16}.
    As a sanity check, we use the quasar spectrum template in \citet{HC16} to fit 
    the {\it WISE} W1 and W2 fluxes of the quasars and define the two subsamples based on 
    the template-estimated rest-frame $2\mu \text{m}$ magnitude. 
    Among all the IR-bright (IR-faint) quasars defined using the power-law fit, 
    11.6\% are classified as IR-faint (IR-bright) in the template-based classification.
    In both cases, IR-bright and IR-faint quasars have similar numbers of faint and intermediate 
    companions (consistent at a $1.5\sigma$ level).
    }
\end{enumerate}

\section{Discussion} \label{sec:discussion}

\subsection{Selection Effects} \label{syserr}

%{\bf I suggest that you mention this earlier in the text, see my previous comments.}
Our sample consists of quasars from the SDSS quasar catalogs and the V\'eron quasar catalog
that have been observed by {\it HST} ACS/WFC broad-band imaging. The parent sample 
(SDSS + V\'eron) includes almost all quasars known to date. 
%The SDSS quasar catalogs 
%are of high completeness. For example, \citet{van05} argued that SDSS quasar selection 
%reached a completeness of ~89\%. 
%Quasars from V\'eron catalog are selected in different ways compared to
%SDSS quasars, and will further enhance the completeness.
Selection effects may be introduced by selecting
quasars that have ACS/WFC imaging.
Among the 532 quasars in our sample, 402 of them (76\%) were observed by ACS/WFC
for purposes that were irrelevant to AGN science (e.g., photometric surveys, 
studies on supernovae or local galaxies). 
%while the quasars we study are in the {\em HST} field by chance
%and will not suffer extra selection effects.
%This fraction is 138 out of 230 (60\%) in sample A and
%248 out of 354 (70\%) in sample B.
Using the subsample of quasars which were observed in AGN-unrelated programs,
we estimate the average companion numbers around quasars to be $0.194\pm0.052$, 
$0.003\pm0.028$ and $0.005\pm0.023$
for faint, intermediate and bright companions.
These numbers are close to the result we get using the whole sample.
For the rest of the sample, the goal of the original {\em HST} 
program was related to AGN science, which 
%The rest of the quasar sample were observed for studies about AGN.
could introduce complicated selection effect, e.g., 
for the programs that targeted a specific class of AGN that could have higher merger fraction.

%{\bf if you only use 461 of them, will the result change?}
%{\bf you didn't discuss: (1) how this might be an selection effect, (2) does it mater? You can use the 461 quasar subsample to answer this question.}
%Future studies with carefully constructed samples
%will be useful to get quantitative results.

Our results described in \S \ref{sec:results} are based on several assumptions,
which may also introduce systematic errors.
The main assumption we make is that the companion fluxes can be 
converted to stellar masses as described in \S \ref{sec:maglim}.
This can not be applied to individual objects. However,
if quasar companions follow the same stellar mass and flux distribution 
as normal galaxies, our method will provide the correct 
numbers of companions that fall in a certain stellar mass range.
This method ignores the possible influence of quasars on their companions,
which is a complicated effect and is difficult to correct.
When applying the results, this potential systematic error must be kept in mind.
%{\bf not clear what this means.}
On the other hand, the ``primary results'' in \S \ref{sec:primary_result}
are free from this systematic error.

\subsection{The Fraction of Merger-Triggered Quasars} \label{alpha}

Our results suggest that there is a deficit of companions around quasars
compared with inactive galaxies.
We interpret the difference as a result of the difference between 
the merging history of quasars (especially merger-triggered ones)
and normal galaxies.
Here we first review the merger-triggering model of quasars briefly
and discuss how many companions would we expect around a 
merger-triggered quasar.

According to simulations \citep[e.g.,][]{lotz10}, the evolution of 
a galaxy merger will experience five stages: the first encounter,
the largest separation, the second encounter, the final coalesce
and the post-merger remnant. Each stage last for $\lesssim 0.5 \text{Gyr}$. 
%{\bf really? the age of the universe at redshift two is only 3 Gyr...}
The typical lifetime of a quasar is $10 \sim 100 \text{Myr}$
\citep[e.g.,][]{hopkins08}, which means that the morphological properties of
quasar host galaxies will not change significantly during the life
of the quasar.

Previous simulations on galaxy major mergers triggering quasars 
\citep[e.g.,][]{springel05, DM05, hopkins06, NK13} 
have suggested that quasars emerge at the final coalesce stage. 
This can be understood in the following picture.
In the merger-triggering model, the quasar activity 
emerges when strong gas inflows feed the SMBH. 
This process requires the gas content to be highly disturbed.
Simulations of galaxy major-mergers show that the disturbed features 
are the most prominent at the second encounter stage.
It takes some time for the gas to reach and feed the SMBH, 
thus strong quasar activities are expected to emerge at the final coalesce.
As an observational evidence, \citet{ellison13} found that 
post-mergers in SDSS have a high AGN-fraction.
%These ``merged" galaxies should have fewer close companions than
%randomly-selected galaxies, since we can only see one galaxy rather than a close pair.
%\textcolor{red}{This argument is also consistent with our result.}

%\textcolor{red}{We use the simulation in \citet{hopkins06} 
%as an example to illustrate this process in details.
%\citet{hopkins06} simulated the merger of two identical disk galaxies
%with virial velocity $V_\text{vir}=160 \text{km/s}$ and initial gas fraction of 20\%.
%The first encounter of the two galaxies happens at about 0.4 Gyr.
%The two galaxies reach their largest separation at about 0.8 Gyr, 
%the second encounter happens at about 1.2 Gyr and the final coalesce is around 1.5 Gyr.
%The AGN activity in this simulation starts at $\sim 1.4$ Gyr and lasted for less than 0.2 Gyr.} 

%{\bf is this a conclusion from the simulations? need refeence.}

With some simple calculations, 
we derive the fraction of merger-triggered quasars
using the number of close companions.
Assuming that quasars are either triggered by secular evolution or mergers,
and the fraction of merger-triggered quasar is $\alpha$.
We use $\overline{N}_\text{comp,QS}$ and $\overline{N}_\text{comp,QM}$
to denote their average number of companions, where ``Q'' means ``quasar'',
``S'' stands for ``secular evolution'' and ``M'' means
``merger''. The observed average companion number of quasars should be 
$\overline{N}_\text{comp,Q}=(1-\alpha)\overline{N}_\text{comp,QS}+\alpha \overline{N}_\text{comp,QM}$,
from which we have
\begin{equation}
\alpha=\frac{\overline{N}_\text{comp,QS}-\overline{N}_\text{comp,Q}}{\overline{N}_\text{comp,QS}-\overline{N}_\text{comp,QM}}
\label{eq:merger_frac}
\end{equation}
We further assume that (1) 
secular-evolution-triggered quasars have the same number of companions as normal galaxies
($\overline{N}_\text{comp,QS}\approx \overline{N}_\text{comp,G}$, where ``G'' means inactive galaxies),
since the secular evolution will not influence the merging process;
and (2) merger-triggered quasars have no close companions
($\overline{N}_\text{comp,QM}\approx 0$),
since the two progenitor galaxies of the merger-triggered quasar 
have already merged into a single galaxy.
We then have
\begin{equation}
\alpha=\frac{\overline{N}_\text{comp,G}-\overline{N}_\text{comp,Q}}{\overline{N}_\text{comp,G}}
\label{eq:merger_frac_obs}
\end{equation}

Figure \ref{fig:density2} indicates that the difference between the average number of companions
of quasars and galaxies varies with the companion stellar mass.
 As a result, $\alpha$ depends on the companion mass.
 This result is not surprising since we expect that major mergers 
 and minor mergers have different probabilities to trigger quasars.
 We can interpret $\alpha(M_\text{companion})$ as the fraction of quasars
 that are triggered by a merger where one progenitor galaxy had
 a stellar mass of $M_\text{companion}$.
 Equation \ref{eq:pfquasar} and \ref{eq:pfgalaxy} describe 
 the average number of companions of our sample.
 According to Equation \ref{eq:merger_frac_obs}, we have $\alpha(\text{faint})=-0.31 \pm 0.26$,
 $\alpha(\text{intermediate})=0.95 \pm 0.25$ 
 and $\alpha(\text{bright})=0.23 \pm 1.14$.
Since intermediate and faint companions are associated with major and minor mergers respectively,
our toy model indicates that there should be a significant fraction of quasars that
are triggered by major mergers, and that minor mergers have a small contribution
in quasar triggering.
The large error for the bright companions does not allow for a strong conclusion.

We now discuss the impact of our assumptions in the analysis above.
The first assumption is that secular-evolution-triggered quasars have 
the same number of companions as normal galaxies. 
Close neighbors of galaxies can disturb their gas kinematics
and lead to gas inflows. 
However, gas inflows generated in this way are usually not strong enough
to feed a quasar, thus the environmental influence should be minor. 
Previous studies have also suggested that secular evolution
is only responsible for faint AGNs \citep[e.g.,][]{treister12}.
Moreover, according to Equation \ref{eq:merger_frac}, 
a larger number of close companions around 
secular-evolution-triggered quasars will lead to a higher 
fraction of merger-triggered quasars.
The second assumption is that there are no companions around merger-triggered quasars.
This assumption is clearly over-simplified. 
Our main idea is that,
%{\bf you need to discuss this a bit more; the immediate reaction from a reader would be all the images of quasars in merger galaxies, including those in this paper! Say something like: this is clearly an over-simplification (observations). Our main pint is that in the picture of a merger-....}
in the picture of a merger-triggered quasar discussed above, 
the quasar host galaxy is a galaxy merger that has entered the final-coalesce stage, 
and we can only see one galaxy rather than a pair. 
%{\bf here you need to discuss the fact that you are using 10kpc cut off, while most of the observed quasar merger system are closer, and the remnants of mergers might still present close by which we are not sensitive to.}
Two possible errors come into this assumption.
On the one hand, a merger-triggered quasar may not show up exactly in the final-coalesce
stage. It is possible that the quasar is triggered earlier
when the two merging galaxies are still distinguishable.
We suggest that the influence of this possible error should be small, however,
because we only count companions with distance larger than 10 kpc,
and most observed merging galaxy pairs are closer.
On the other hand, 
if there are nearby galaxies that are not involved in this merging event,
we will have a non-zero companion number for merger-triggered quasars.
Given the distance cut we applied $(10\text{ kpc}<d<30\text{ kpc})$,
such cases should be rare.
In both cases, Equation \ref{eq:merger_frac} suggests that, 
if $\bar{N}_\text{comp, QM}>0$, 
the fraction of merger-triggered quasars will become even larger,
which will further strengthen our conclusion.

\subsection{A Unified Picture of AGN Triggering}\label{sec:triggermodel}
Conclusions from previous studies on AGN merger fractions have been ambiguous.
There are both results indicating enhanced merging fractions of AGN
\citep[e.g., ][]{ellison11, silverman11, satyapal14, fan16, weston17, goulding17}
and indicating no difference between AGN and normal galaxies
\citep[e.g., ][]{cisternas11, schawinski11, kocevski12, villforth17}.
Our result suggests that there should be a significant fraction of 
major-merger-triggered quasars. We consider several possible reasons
for this ambiguity.

%The reason of this ambiguity could be complicated.
Firstly, most of these studies used distinct AGN samples.
The AGN samples vary from emission-line-ratio selected
\citep[e.g.,][]{ellison11}, 
near-IR selected \citep[e.g.,][]{satyapal14, fan16, weston17, goulding17},
X-ray selected \citep[e.g.,][]{cisternas11, silverman11, kocevski12, villforth17}, 
to optical selected (this work).
We notice that all the studies using near-IR selected AGN
samples reach a conclusion that AGN have an enhanced 
merging fraction, while most of the X-ray selected AGN samples
 do not show a significant difference from inactive galaxies.
This result can be explained if different populations of AGN 
emerge in different stages of galaxy mergers.
%This point will be discussed in details in Section \ref{mergermodel}.
The luminosities of the AGN samples are also different,
which is believed to have a crucial influence on AGN merging fraction.
Previous observations using optical data focused mainly on 
low-luminosity AGNs to avoid the strong emission from the central nuclei. 
Most studies on high-luminosity AGNs are based on 
either X-ray or infrared data (for obscured ones).
Comparing to these studies, 
our sample consists of optical-selected AGNs and 
spans a wide range of luminosity $(-31<M_i(z=2)<-23)$.
%{\bf mention the luminosity range of our sample compared to others.}

Secondly, the method of these studies might introduce some biases.
Most of the previous studies used disturbed features in quasar host galaxies
as indicators of recent merger events.
%However, the disturbed features
%might have faded away.
% {\bf avoid using words such as inevitable problem; this paper will be refereed by someone who did these previous work, and you don't want to appear to criticize their work unless you have to. }
There have been arguments that these features should be able to survive until 
the quasar activity emerges \citep[e.g.,][]{cisternas11, villforth17}.
However, this depends on the model of galaxy mergers and AGN triggering,
which is highly uncertain and varies from object to object.
Even if the average timescale of disturbed features may be long enough, 
it is still possible to miss some mergers and thus underestimate
the merger fraction of AGN.
The uncertainty in identifying disturbed features in AGN host galaxies
might be more severe for bright unobscured AGNs that need PSF subtraction,
given the difficulty of PSF modeling.  

It is also difficult to distinguish 
major mergers and minor mergers based on the host galaxy morphology.
If minor mergers do not contribute to the triggering of quasars (as indicated by our result),
including them as ``recent merging systems'' will increase the number of identified
merging systems and introduce extra uncertainties 
when testing the major-merger-triggering mechanism.

%Compared to the traditional method which looks for 
%disrupted features in quasar host galaxies,
%our analysis has some important advantages.
In comparison, we identify mergers by counting companions. 
This  will not introduce significant biases
against the late-stage mergers, where the disturbed features in the
galaxies might have already faded away.
It is also more straightforward to distinguish major
and minor mergers since we can estimate the companion mass.
Counting companions is an accessible way for nearly all 
populations of AGN, and does not require superb angular resolution,
which makes it easier to build a large, unbiased sample.

Keeping the possible biases in mind,
we find that most of the results are consistent with the picture where:

(1) Mergers are the main triggering mechanism for high-luminosity AGNs, and secular evolution is mainly 
responsible for low-luminosity AGNs. There have been simulations claiming that 
secular evolution is not powerful enough to trigger the most luminous quasars \citep[e.g.,][]{treister12}.
Previous observations have also reported that the merger fraction increases with AGN luminosity
\citep[e.g.,][]{fan16}.
We find that the average number of intermediate companions decreases with quasar luminosity.
According to Figure \ref{fig:pf_m} and Equation \ref{eq:merger_frac_obs}, 
our result indicates that luminous quasars are more likely to be triggered by major mergers.

 (2) Merger-triggered AGNs evolve from an obscured to an unobscured phase.
 The transition happens when the radiation and 
material outflows blow away the dust around the active nucleus.
Consequently, IR-selected AGNs (which are dustier) might represent an earlier stage of 
AGN evolution, and it is easier to detect the disturbed features in their host galaxies
than other populations of AGN.
This picture is supported by simulations
\citep[e.g.,][]{DM05,hopkins08}, and can explain the discrepancy with previous observations,
which reported that IR-selected AGNs have a larger merging fraction than X-ray-selected AGNs.

\subsubsection{Merger Rate of AGN vs AGN Rate in Mergers}
%In addition to the difference in the AGN sample and the way to identify mergers,
%it might also be confusing how to quantify the importance of major mergers in AGN triggering.
In comparison to  works such as ours,  which measures the fraction of mergers in quasars,
some studies instead compared the fraction of AGN in merging and non-merging
systems, quantified by the ``AGN fraction ratio'':
\begin{equation} \label{eq:R}
R=\frac{P(\text{AGN}|\text{merger})}{P(\text{AGN}|\text{non-merger})}
\end{equation}
where $P(\text{AGN}|\text{merger})$ ($P(\text{AGN}|\text{non-merger})$) 
is the probability of finding an AGN in a (non-)merging system.
Previous studies found that merging systems are more likely to host AGN
(e.g., $R\sim2-7$ in \citet{goulding17}, $R\sim5-17$ in \citet{weston17}),
and concluded that galaxy mergers are the dominant triggering 
mechanism of the AGN in their sample.
%A large $R$ value (e.g., $R\sim5$) might lead to a conclusion that 
%major mergers dominate the AGN triggering, as suggested in some of the 
%previous studies.
Using Bayesian analysis, we can calculate the AGN fraction ratio using the 
merger ratio in AGN, and thus compare our result directly with previous studies.
According to Bayes' Theorem,
%\begin{equation}
\begin{multline}
P(\text{AGN}|\text{merger})=\frac{P(\text{merger}|\text{AGN})P(\text{AGN})}{P(\text{merger})}\\
P(\text{AGN}|\text{non-merger})=\frac{P(\text{non-merger}|\text{AGN})P(\text{AGN})}{P(\text{non-merger})}
\end{multline}
thus
\begin{equation}\label{eq:R_final}
\begin{split}
R &=\frac{P(\text{non-merger})}{P(\text{merger})}\times\frac{P(\text{merger}|\text{AGN})}{P(\text{non-merger}|\text{AGN})}\\
 &=\frac{1-P(\text{merger})}{P(\text{merger})}\times\frac{P(\text{merger}|\text{AGN})}{1-P(\text{merger}|\text{AGN})}
\end{split}
\end{equation}
where we apply $P(\text{merger})+P(\text{non-merger})=1$.

%Following the studies mentioned above, we use disturbed host galaxy as indicators of mergers.
In our sample, the estimated major-merger-triggered quasar fraction is $\alpha = 0.95 \pm 0.25$,
with a $3\sigma$ lower limit of 0.22.
We assume that the secular-evolution-triggered quasars have the same merger fraction as
inactive galaxies (i.e., $P_\text{S}(\text{merger}|\text{AGN})=P(\text{merger})$).
%and that all the major-merger-triggered quasars have disturbed host galaxies
We also have $P_\text{M}(\text{merger}|\text{AGN})=1$ by definition.
We adopt the merger fraction of inactive galaxies from \citet{villforth17},
which gave $P(\text{merger})\approx 0.2$.
This gives 
\begin{equation}
\begin{split}
P(\text{merger}|\text{AGN}) & =\alpha P_\text{M}(\text{merger}|\text{AGN})+\\
& ~~~~(1-\alpha)P_\text{S}(\text{merger}|\text{AGN})\\
& \geq 0.37,\\
\end{split}
\end{equation}
and $R\geq 2.4$ according to Equation \ref{eq:R_final},
which is consistent with previous studies.

%\subsection{Constraining The Major-Merger-Triggering Models of Quasars} \label{mergermodel}

%\textcolor{red}{Add a bit discussion about thoughts of current community.}

\section{Summary} \label{sec:sum}
We investigate the numbers of companions around quasars
which have $M_i(z=2)<-23$ at $0.3<z<3$.
Based on the SDSS quasar catalogs and the V\'eron quasar catalog,
we construct a sample of 532 quasars
which have been observed by {\em HST} ACS/WFC,
and use the archival images to find all the companions around
these quasars with projected distance of $10\text{ kpc}<d<100\text{ kpc}$.
PSF subtraction is done for all the quasars to enhance
the detectability of close companions,
and the fraction of missed companions was estimated to be less than 10\%
even for the faintest companion at the smallest projected distance of interest
(\S \ref{sec:completeness} and \S \ref{sec:maglim}).
We use galaxies in the 3D-HST photometric catalog to construct 
a redshift-matched sample of massive inactive galaxies
as our control sample.
We define ``faint'', ``intermediate'' and ``bright'' companions such that
faint and bright companions are associated with minor mergers,
and intermediate companions correspond to major mergers.
 We calculate the average number of companions of quasars and inactive galaxies,
 and raise an explanation to the difference between 
 quasars and normal galaxies.
 Our main conclusions are:
 \begin{enumerate}
 \item{Both quasars and inactive galaxies show excesses
 of companion surface densities in their neighborhoods.
 Quasars show a deficit of intermediate companions at projected distance $d\lesssim60$ kpc,
 and have numbers of faint/bright companions similar to normal galaxies.
 The average number of companions around quasars show little evolution with redshift,
  and do not show significant dependence on 
 absorption features, radio loudness and IR luminosity. 
 More luminous quasars have more faint companions and fewer intermediate companions,
 though both trends are not significant.
 The number of bright companions does not evolve with quasar luminosity.
 }
 
 \item{By assuming that merger-triggered quasars have no close companions
 and secular-evolution-triggered quasars have the same number of companions
 as inactive galaxies, the deficit of close companions around quasars
 indicates that a significant fraction of quasars are triggered by major mergers.}
 
 \item{Most of the previous studies are consistent with the picture
 where the merger-triggered fraction increases with AGN luminosity, and 
 merger-triggered AGN evolves from an obscured to an 
 unobscured phase. The ambiguity of previous results may be 
 a result of biases introduced by the samples and the methods.}
 \end{enumerate}

 Using close companions as identifiers of merging systems
 does not require superb angular resolution,
 which makes it possible to constrain the AGN merger fraction using 
 ground-based imaging.
 Ground-based surveys like the Hyper Suprime-Cam Survey
 \citep{aih18} and the Large Synoptic Survey Telescope \citep{lsst}
 have angular resolution that is good enough for companion counting,
 and can provide a large sample to decrease the statistical error.
 Future studies utilizing these surveys will provide a more accurate
 estimate on the fraction of merger-triggered AGN,
 and put better constraints on AGN triggering models.

\acknowledgements
We acknowledge the support from HST-AR-14312 grant from
the Space Telescope Science Institute.
We acknowledge the useful suggestions from George H. Rieke, Ann Zabludoff,
Richard Green and Peter S. Behroozi.
We acknowledge the useful comments and suggestions from the referee.

Based on observations made with the NASA/ESA Hubble Space Telescope, 
and obtained from the Hubble Legacy Archive, which is a collaboration 
between the Space Telescope Science Institute (STScI/NASA), 
the Space Telescope European Coordinating Facility (ST-ECF/ESA) 
and the Canadian Astronomy Data Centre (CADC/NRC/CSA).

This work is based on observations taken by the 3D-HST Treasury Program 
(HST-GO-12177 and HST-GO-12328) with the NASA/ESA Hubble Space Telescope, 
which is operated by the Association of Universities for Research in Astronomy, Inc., 
under NASA contract NAS5-26555.

\facilities{HST}

\appendix
\section{On-Going Merging Systems} \label{ap1}
\begin{figure*}
\gridline{\fig{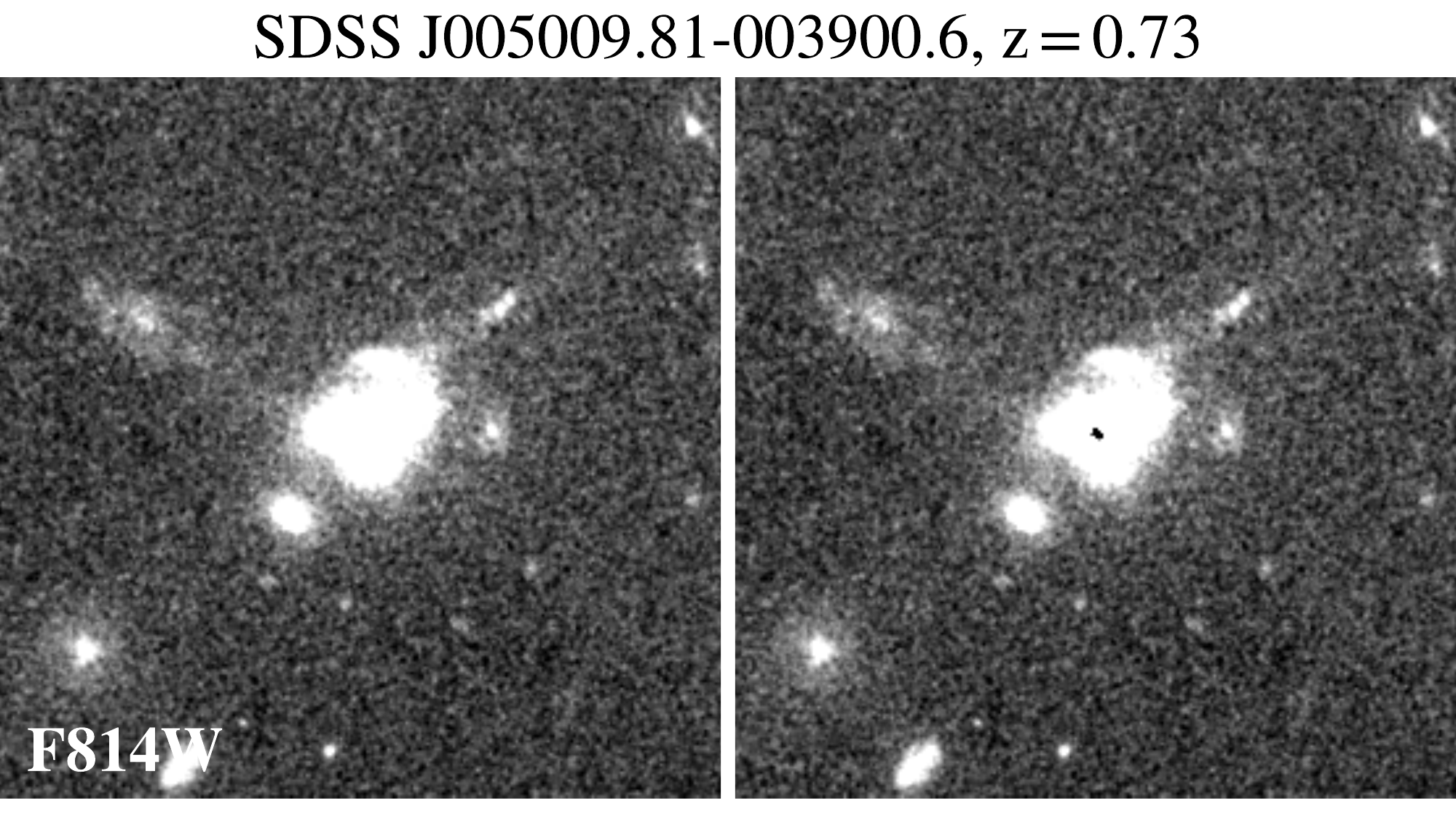}{0.45\textwidth}{}
          \fig{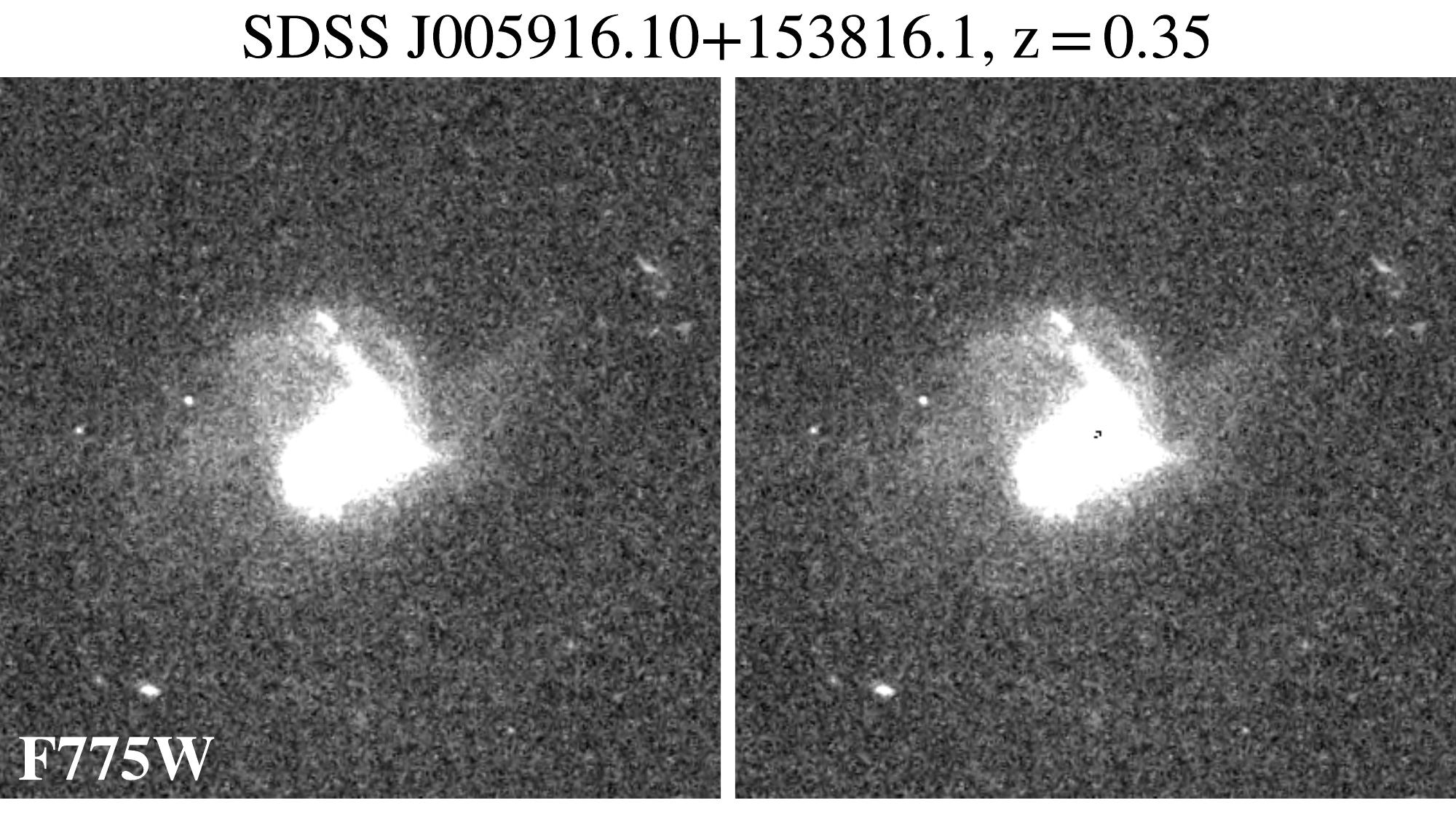}{0.45\textwidth}{}}
\gridline{\fig{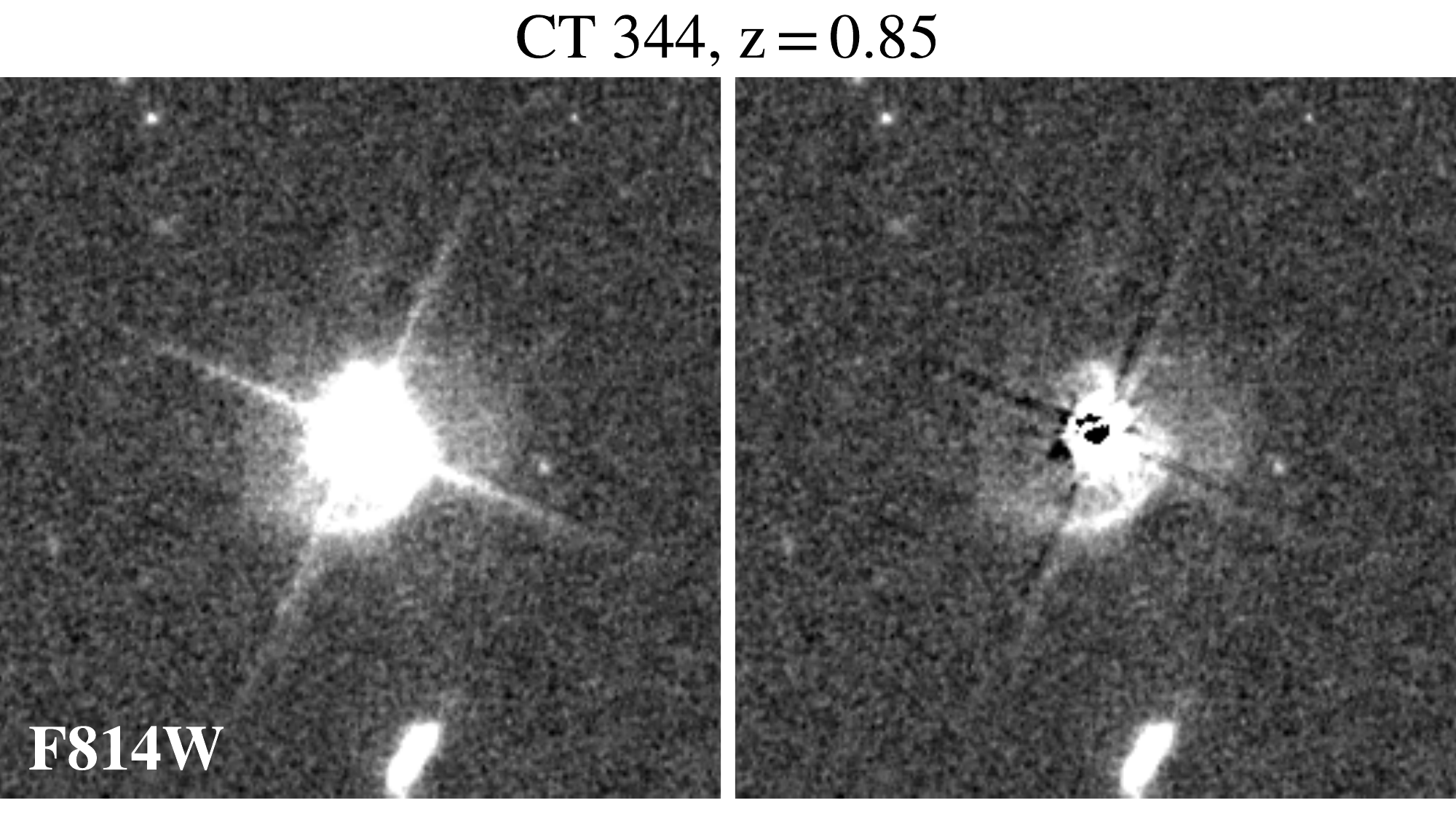}{0.45\textwidth}{}
          \fig{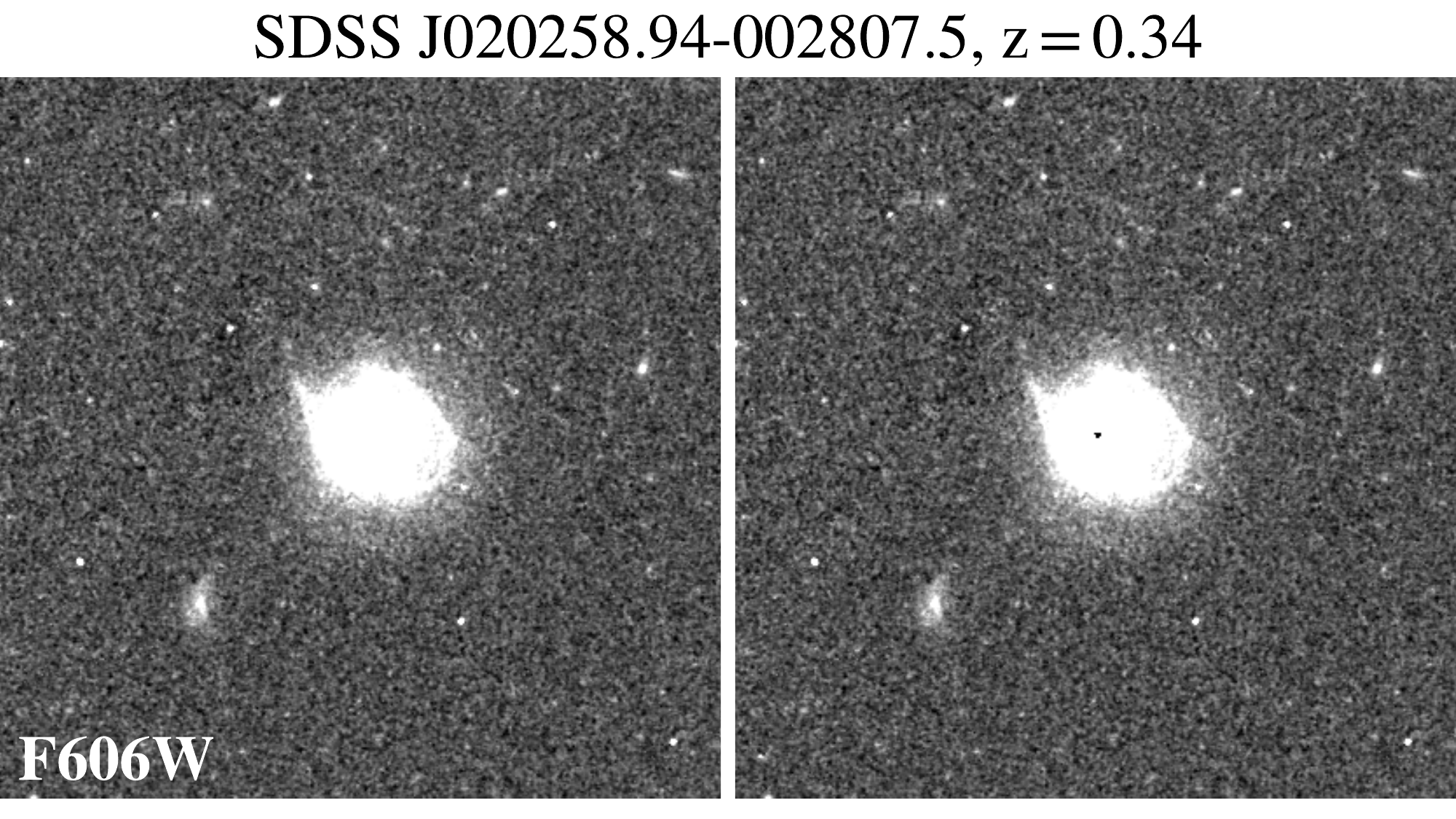}{0.45\textwidth}{}}
\gridline{\fig{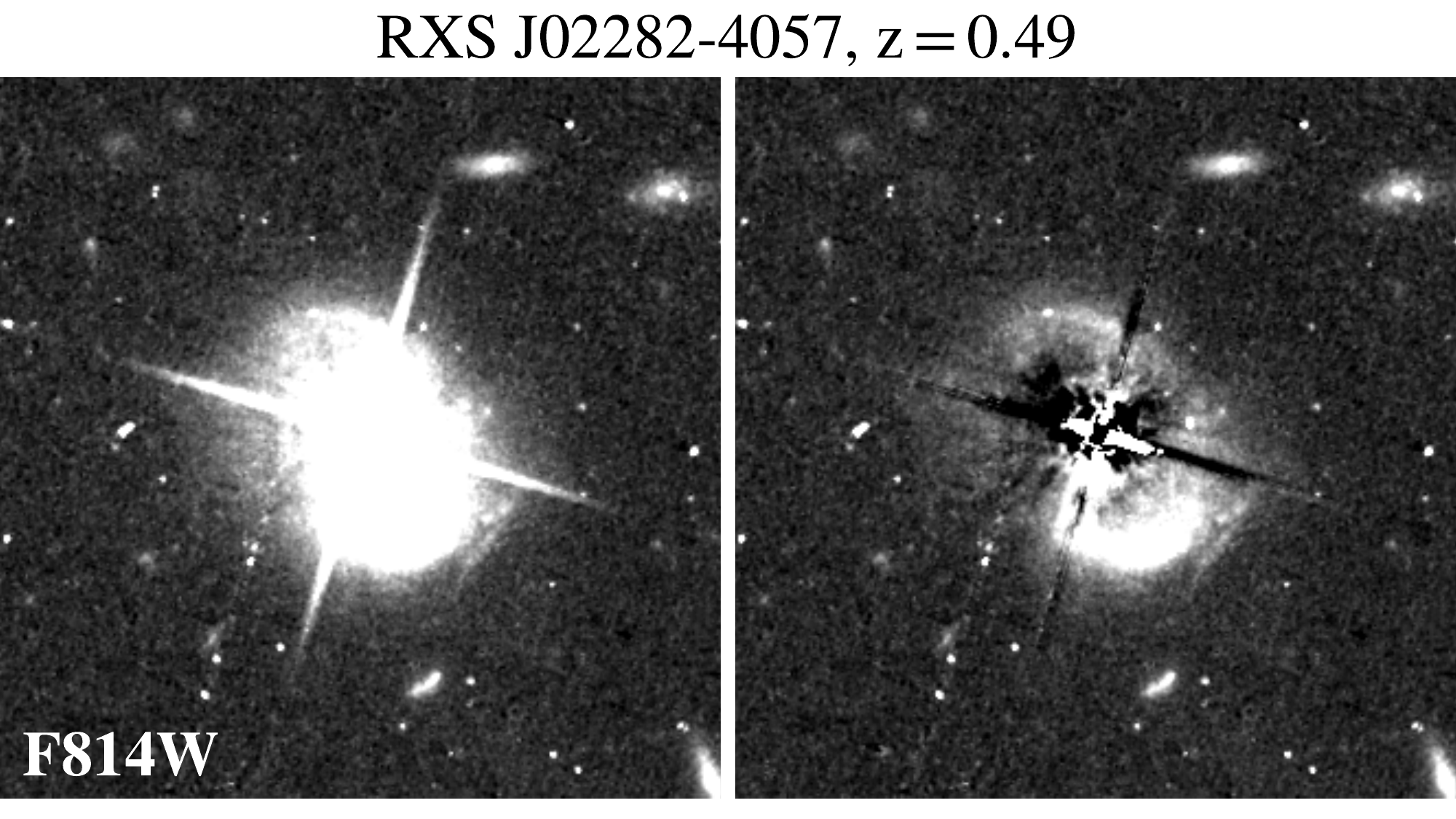}{0.45\textwidth}{}
		\fig{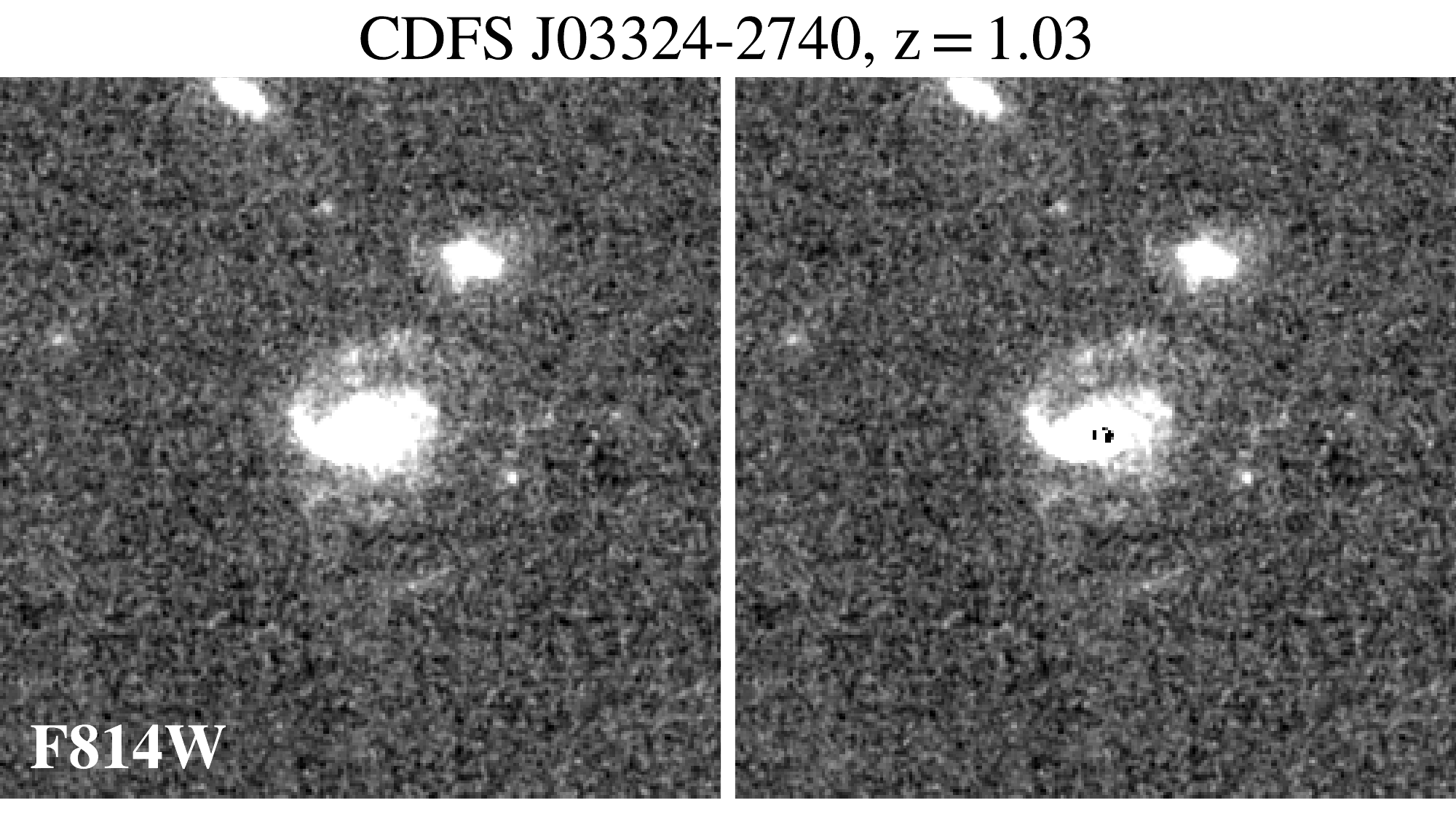}{0.45\textwidth}{}}
\gridline{\fig{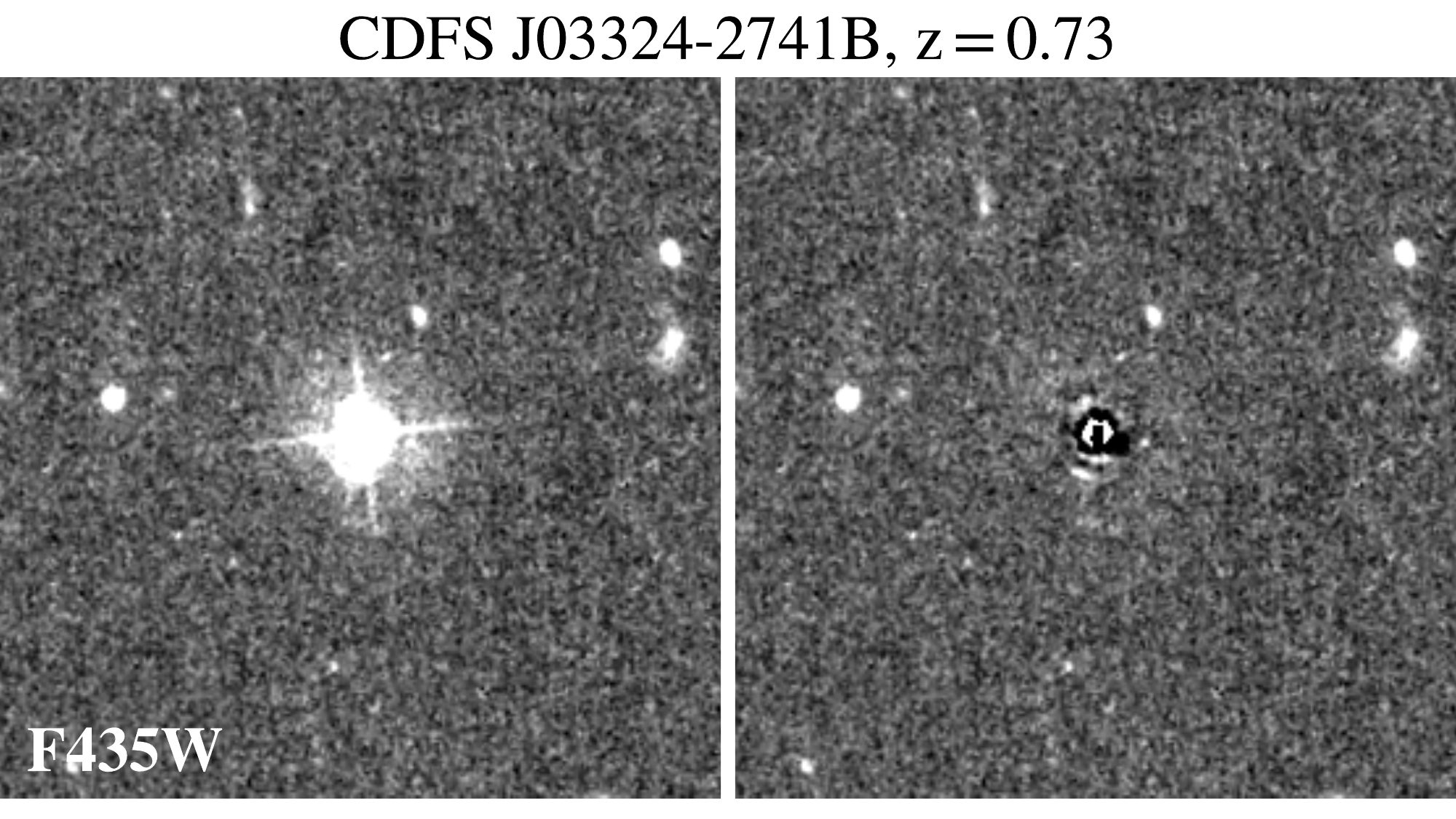}{0.45\textwidth}{}
		\fig{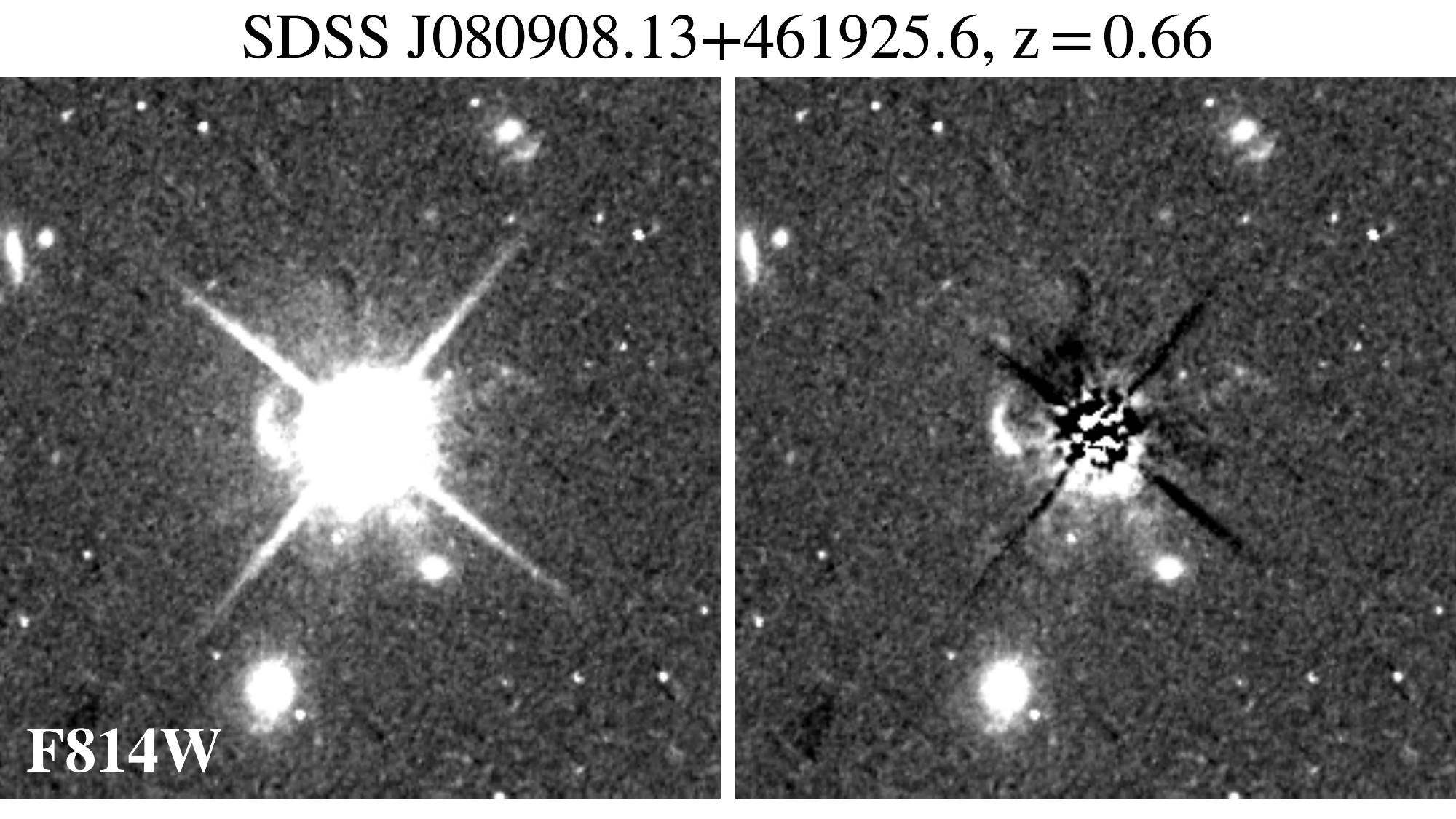}{0.45\textwidth}{}}
\caption{Images of on-going merging systems in our master sample. 
The image sizes are $100\text{kpc}\times 100$kpc at the redshift
of the object. We present both the original image (left) and the 
PSF-subtracted image (right).\label{fig:merger}}
\end{figure*}

In this appendix, we present the images and information of 
candidates of on-going merging systems with quasar activity
mentioned in \S \ref{subsec:candidates}.
We visually inspected all the quasars in the master sample 
and select systems that show either a pair of interacting galaxies 
or some disturbed features like tidal tails.
Figure \ref{fig:merger} shows the images of the on-going merging systems.
All the images are 100 kpc $\times$ 100 kpc in size.
The information of these objects can be found in Table \ref{tbl:merger}.

\begin{deluxetable*}{c|cccc}
\tablecaption{Information of On-Going Merging Systems.\label{tbl:merger}}
\tablehead{
\colhead{Quasar Name} & \colhead{RA} & \colhead{DEC} 
& \colhead{Redshift} & \colhead{Feature}}
\startdata
SDSS J005009.81-003900.6 & 00:50:09.81 &-00:39:00.6 & 0.728 & Interacting Galaxies \\
SDSS J005916.10+153816.1 & 00:59:16.10 &+15:38:16.1 & 0.354 & Interacting Galaxies \\
SDSS J020258.94-002807.5 & 02:02:58.94 &-00:28:07.5 & 0.339 & Tidal Tail \\
SDSS J080908.13+461925.6 & 08:09:08.13 &+46:19:25.6 & 0.657 & Tidal Tail \\
SDSS J110556.18+031243.1 & 11:05:56.18 &+03:12:43.1 & 0.353 & Interacting Galaxies \\
\enddata
\tablecomments{Table 3 is published in its entirety in the electronic 
edition of the {\it Astrophysical Journal}.  A portion is shown here 
for guidance regarding its form and content. The full table is also available
in FITS format at \url{https://github.com/yuemh/qso_companion}.}
\end{deluxetable*}

\renewcommand{\thefigure}{\arabic{figure} (Continued)}
\addtocounter{figure}{-1}

\begin{figure*}
\gridline{\fig{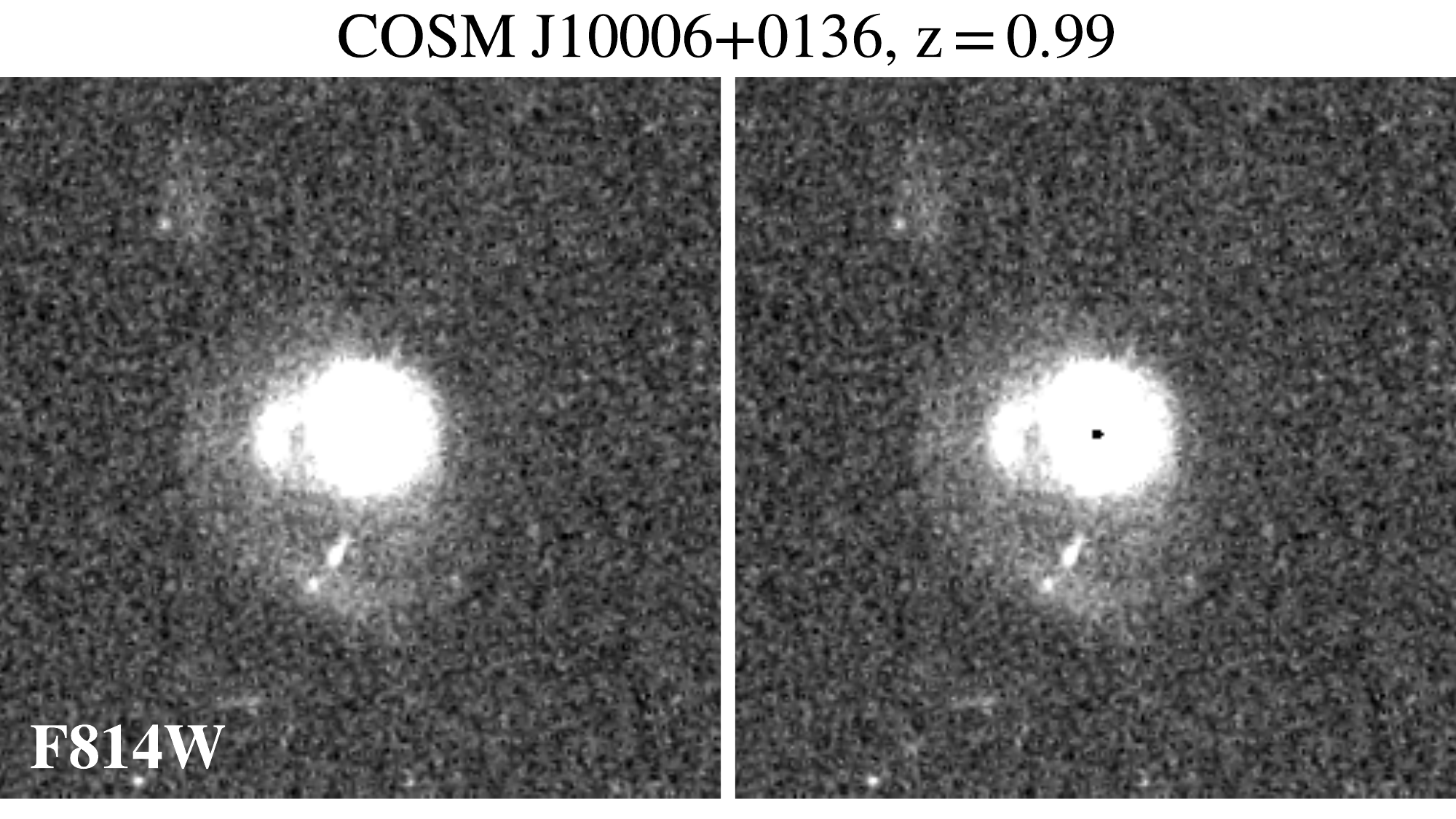}{0.45\textwidth}{}
		\fig{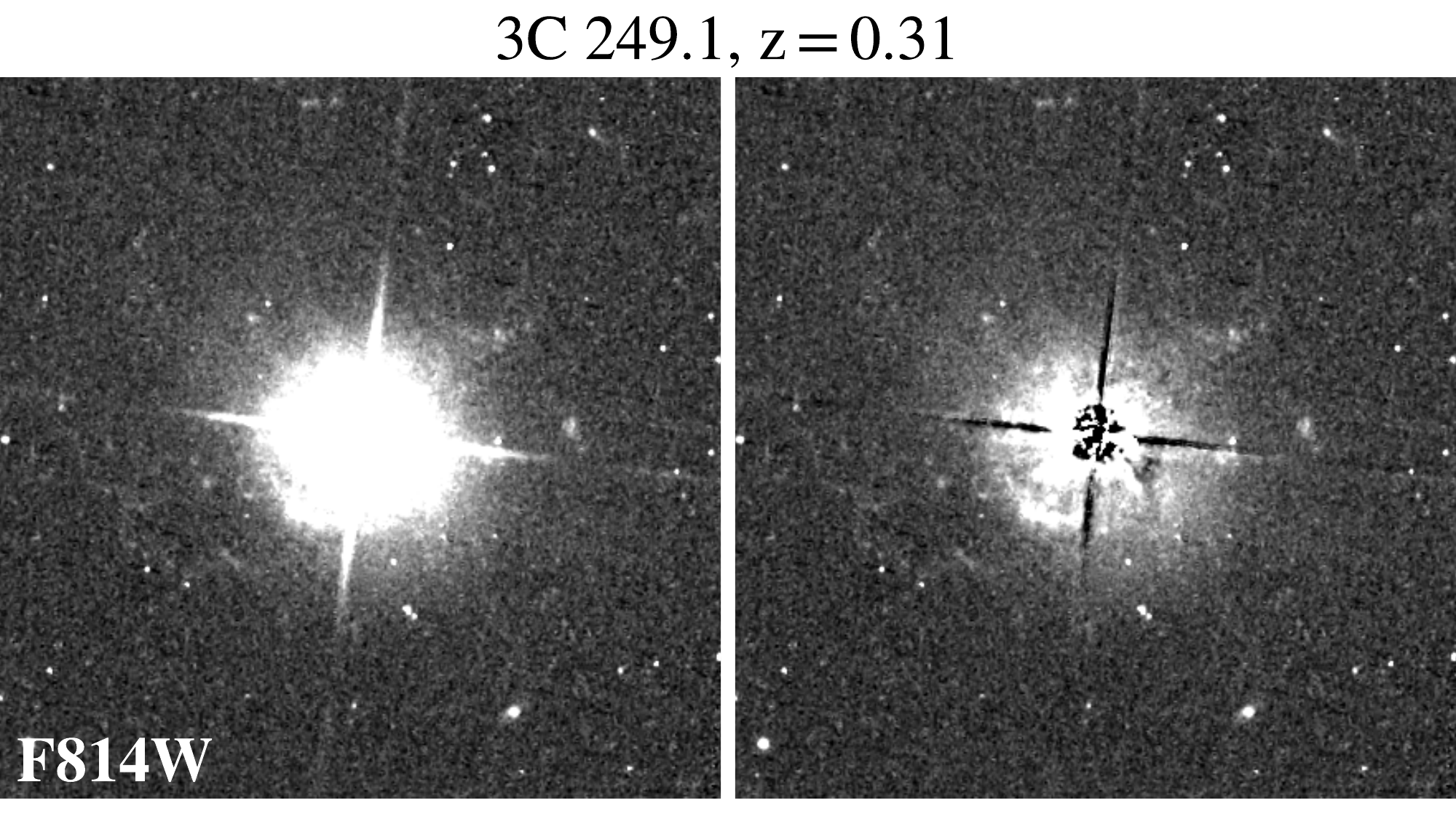}{0.45\textwidth}{}}
\gridline{\fig{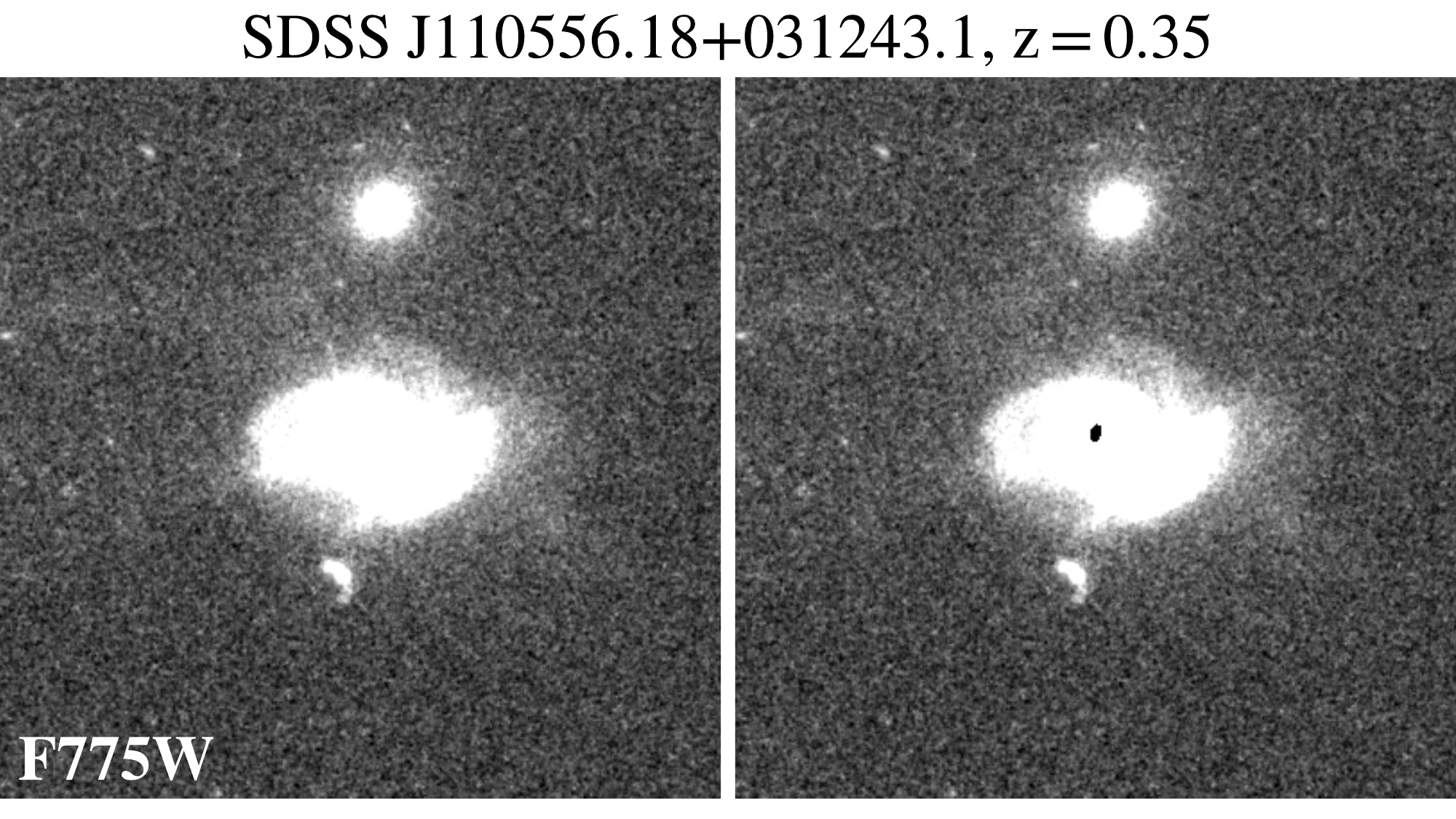}{0.45\textwidth}{}
		\fig{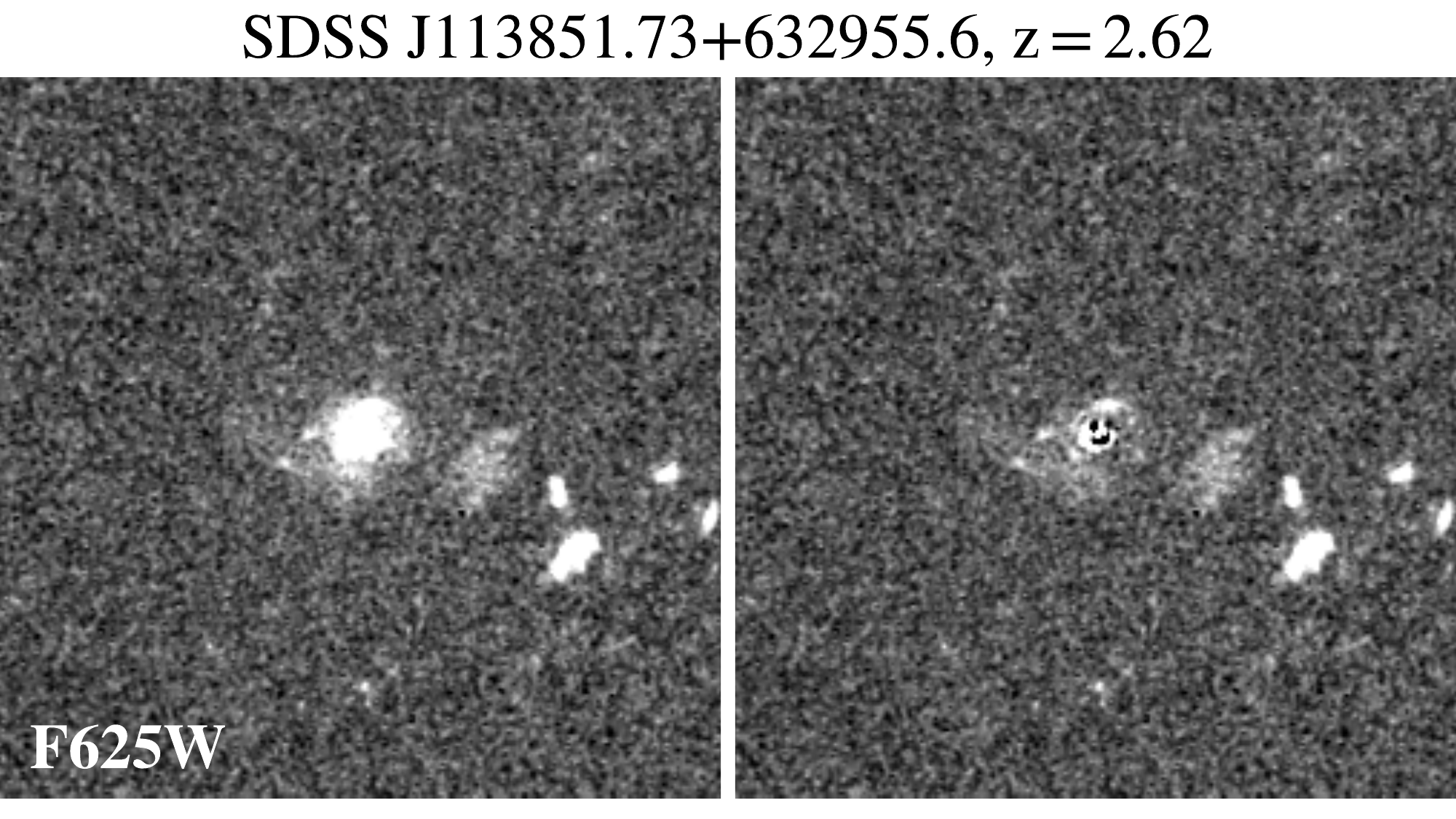}{0.45\textwidth}{}}
\gridline{\fig{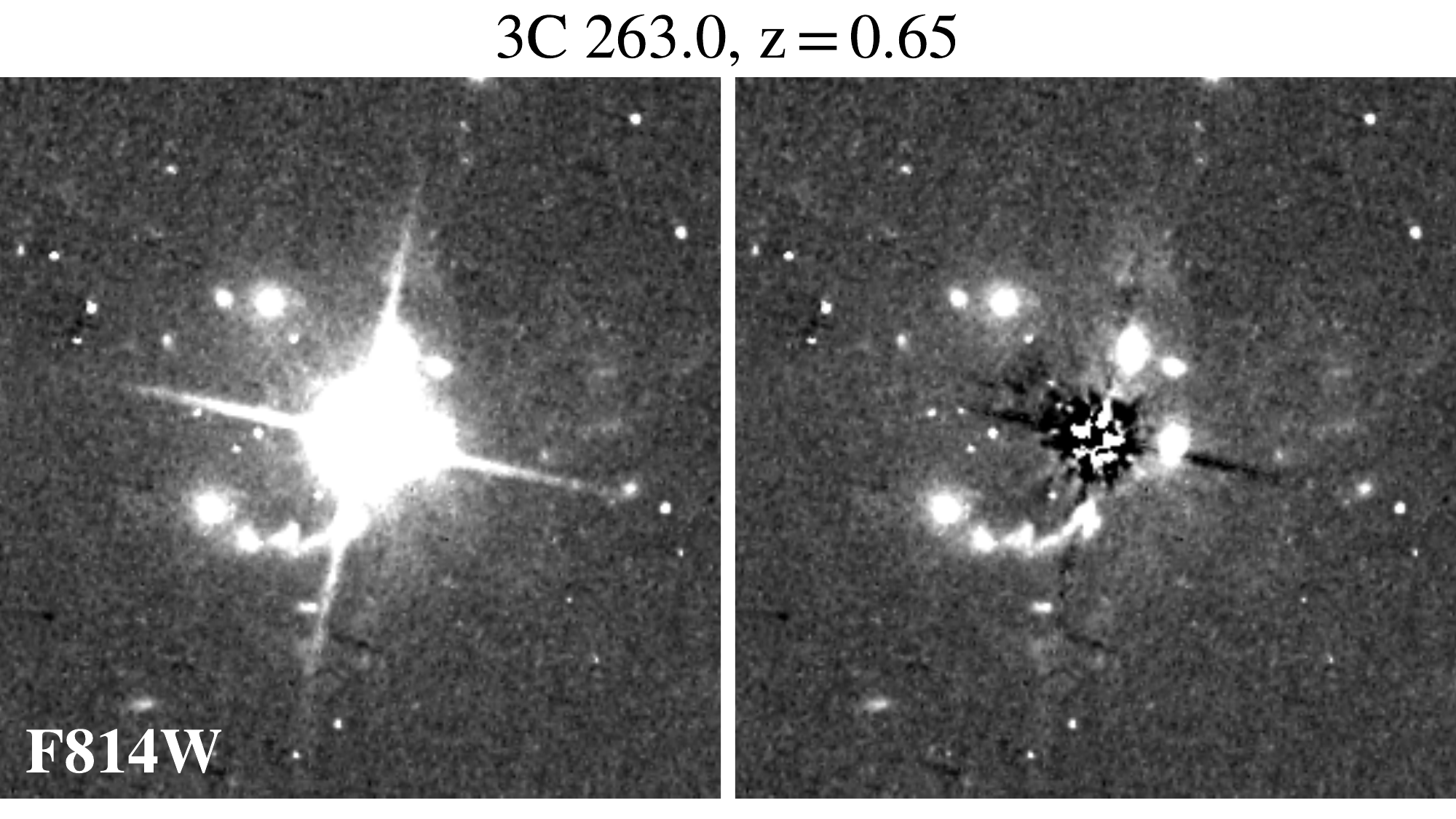}{0.45\textwidth}{}
		\fig{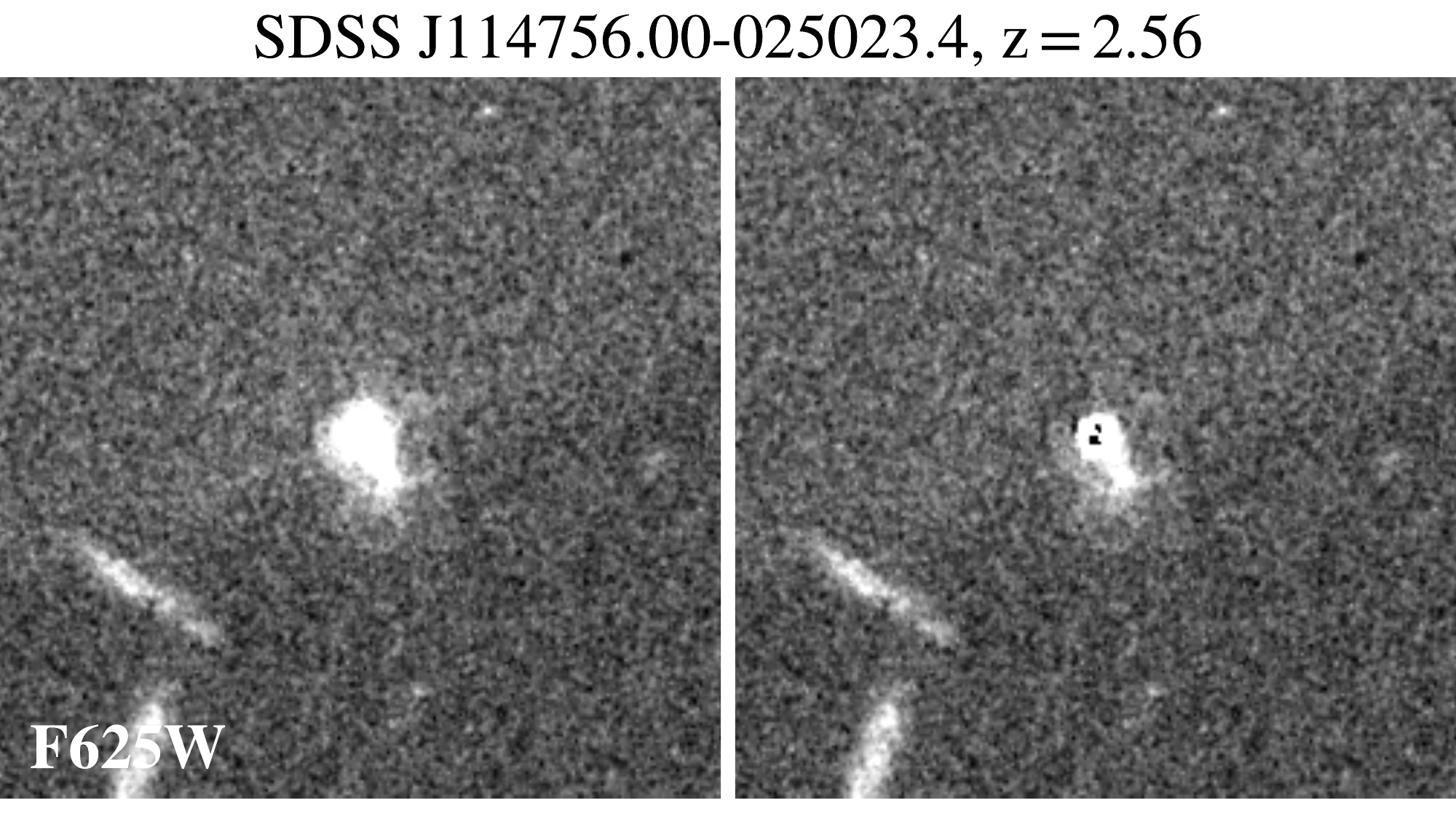}{0.45\textwidth}{}}
\gridline{\fig{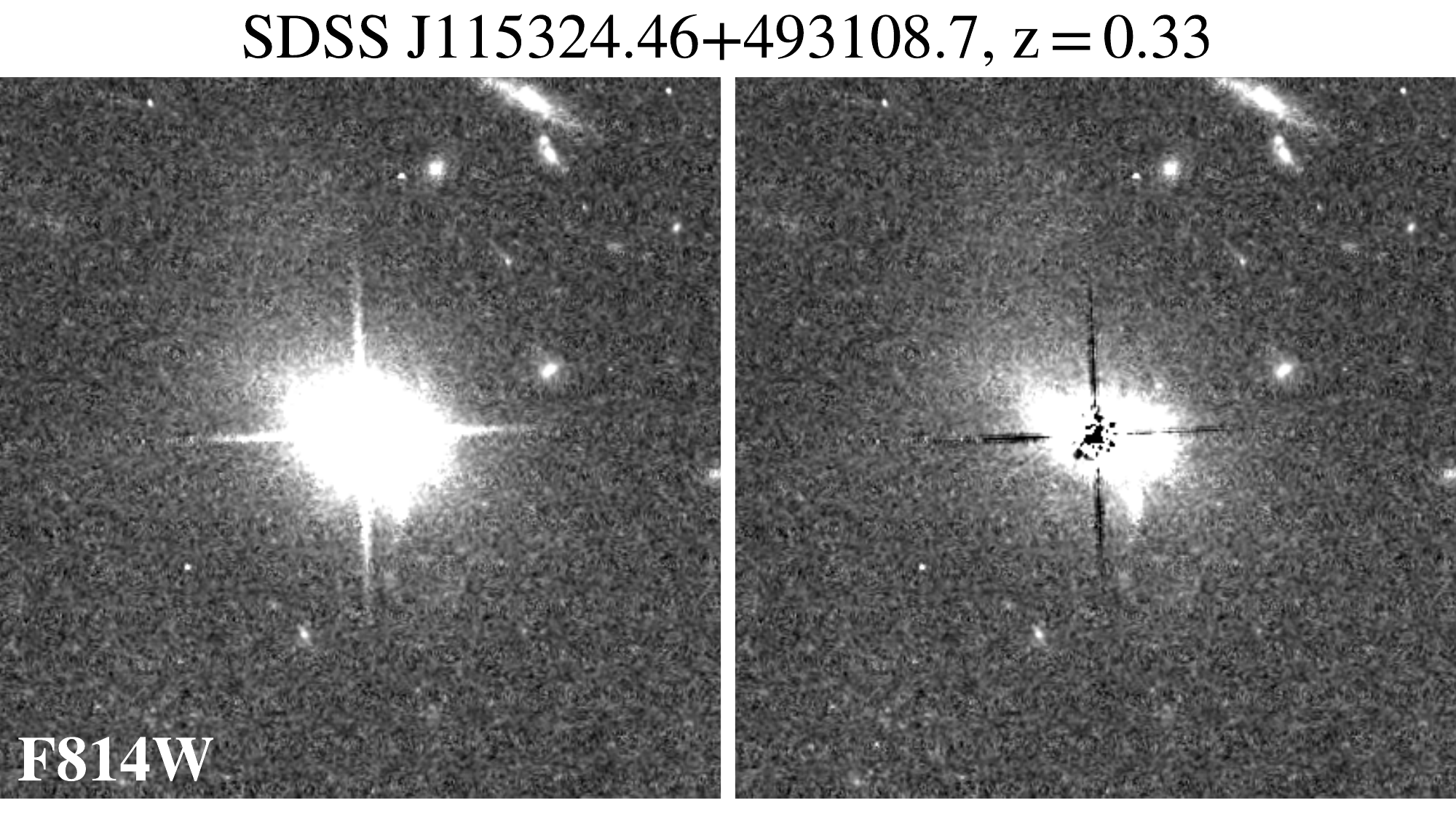}{0.45\textwidth}{}
		\fig{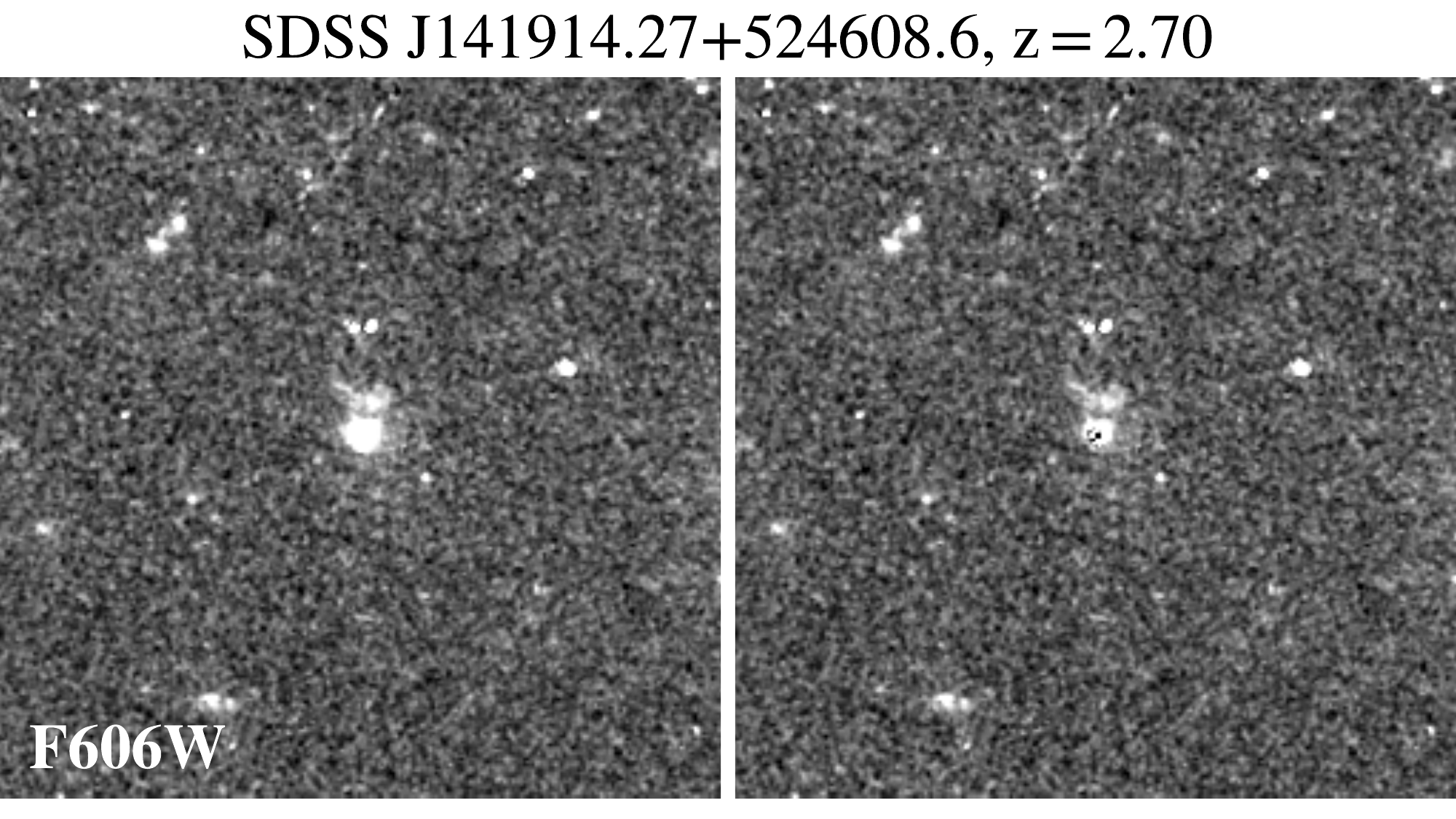}{0.45\textwidth}{}}
\caption{}
\end{figure*}
\renewcommand{\thefigure}{\arabic{figure}}

\renewcommand{\thefigure}{\arabic{figure} (Continued)}
\addtocounter{figure}{-1}

\begin{figure*}
\gridline{\fig{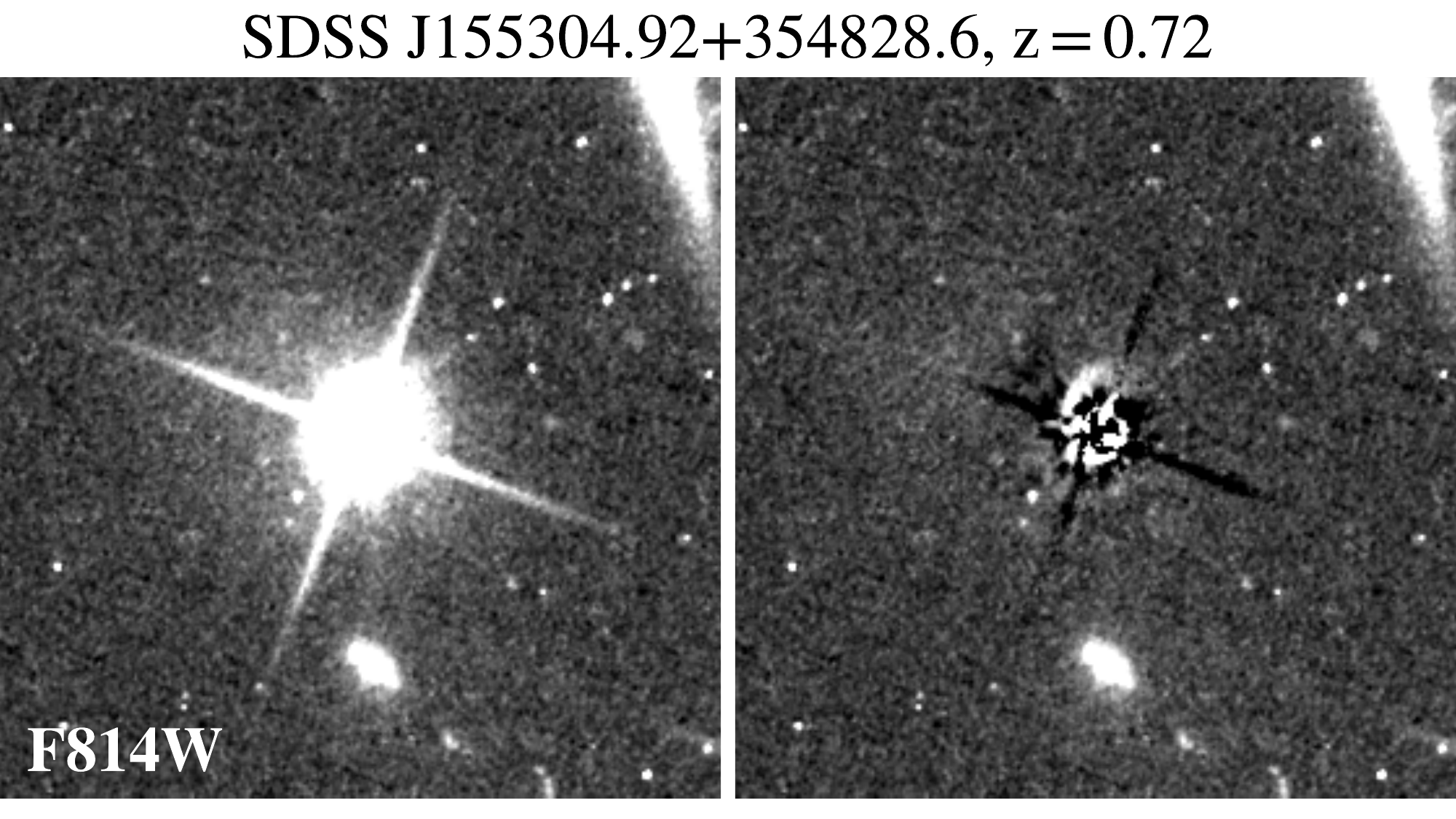}{0.45\textwidth}{}
		\fig{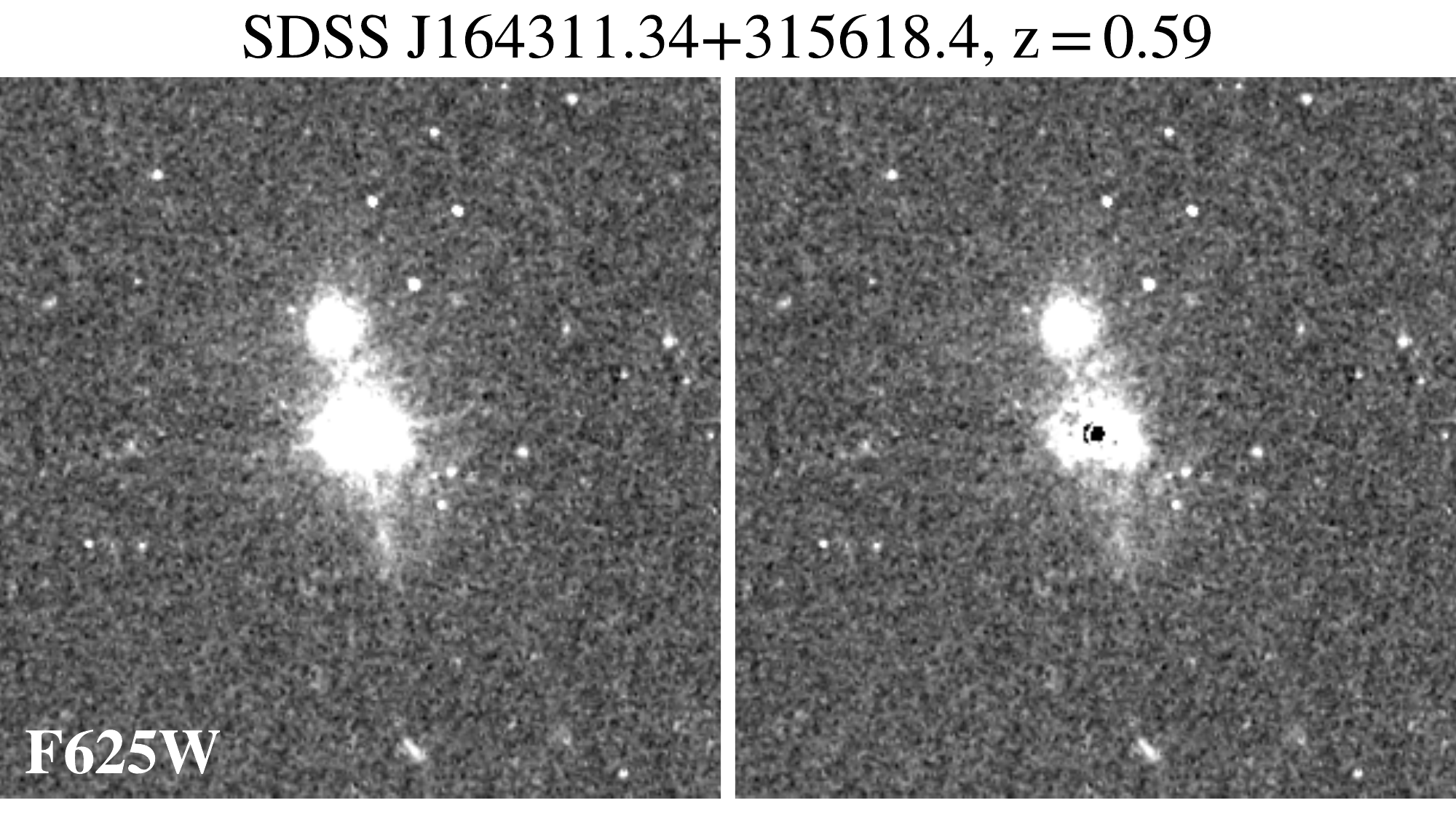}{0.45\textwidth}{}}
\gridline{\fig{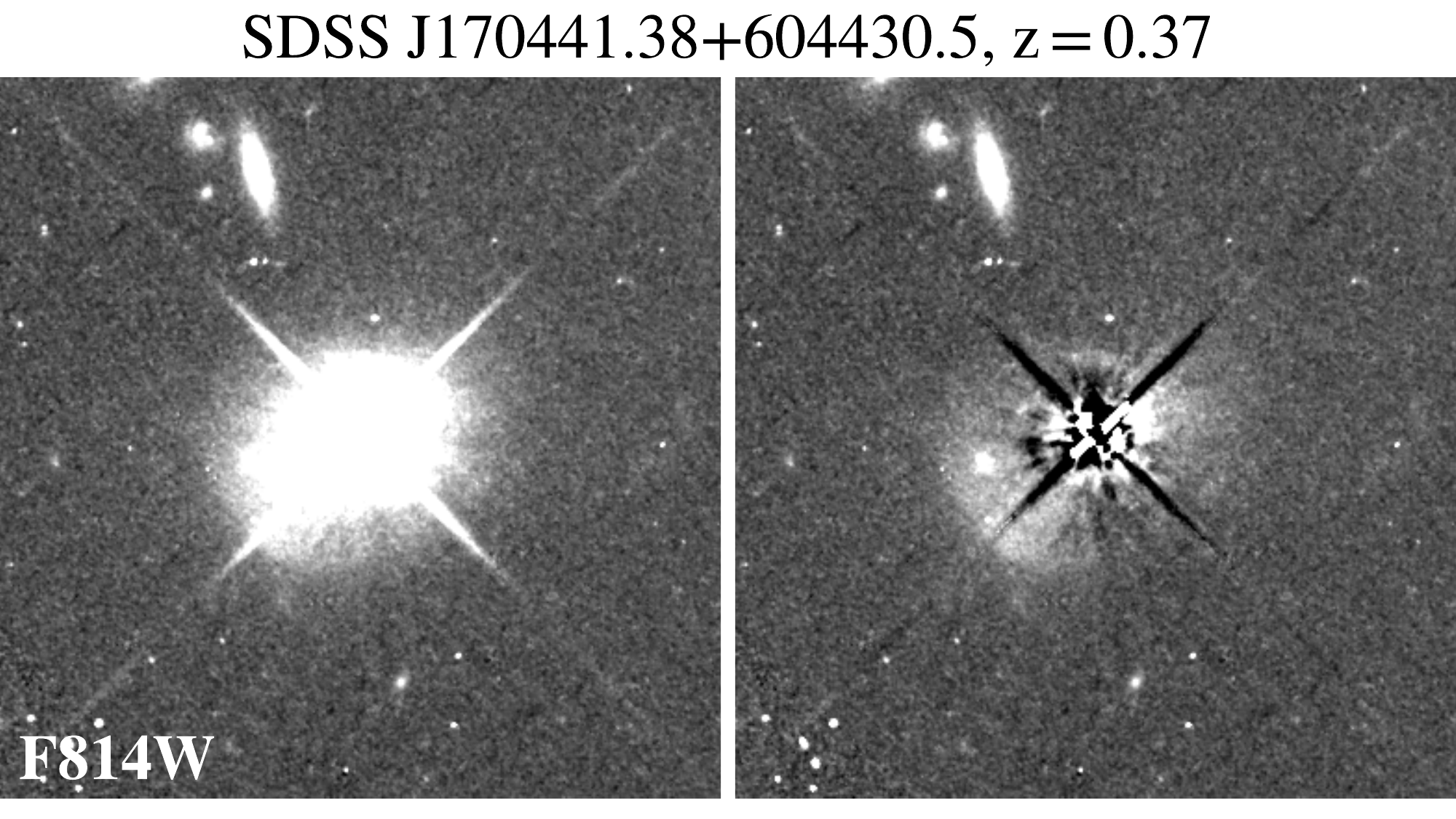}{0.45\textwidth}{}
		\fig{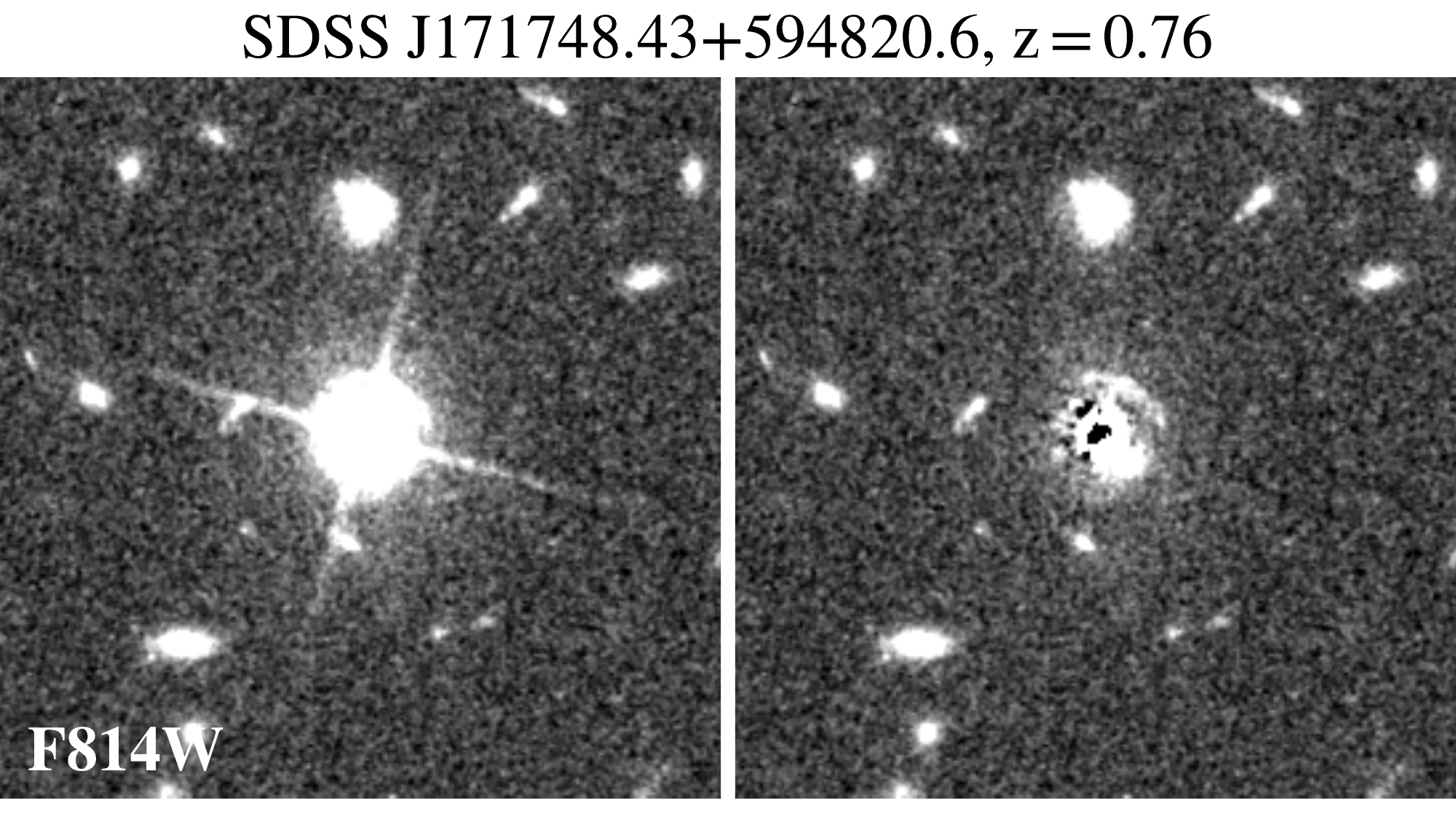}{0.45\textwidth}{}}
\gridline{\fig{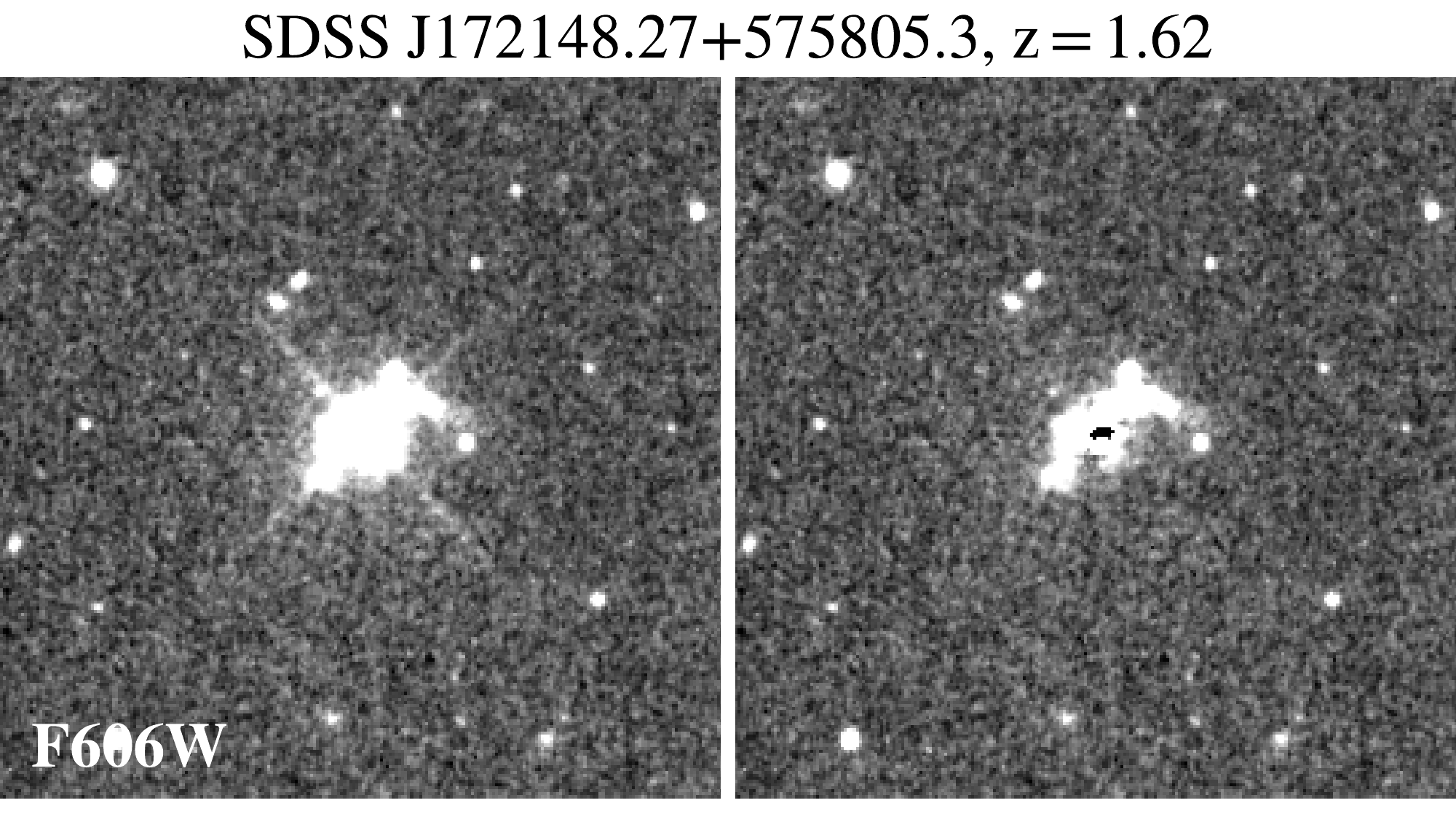}{0.45\textwidth}{}
		\fig{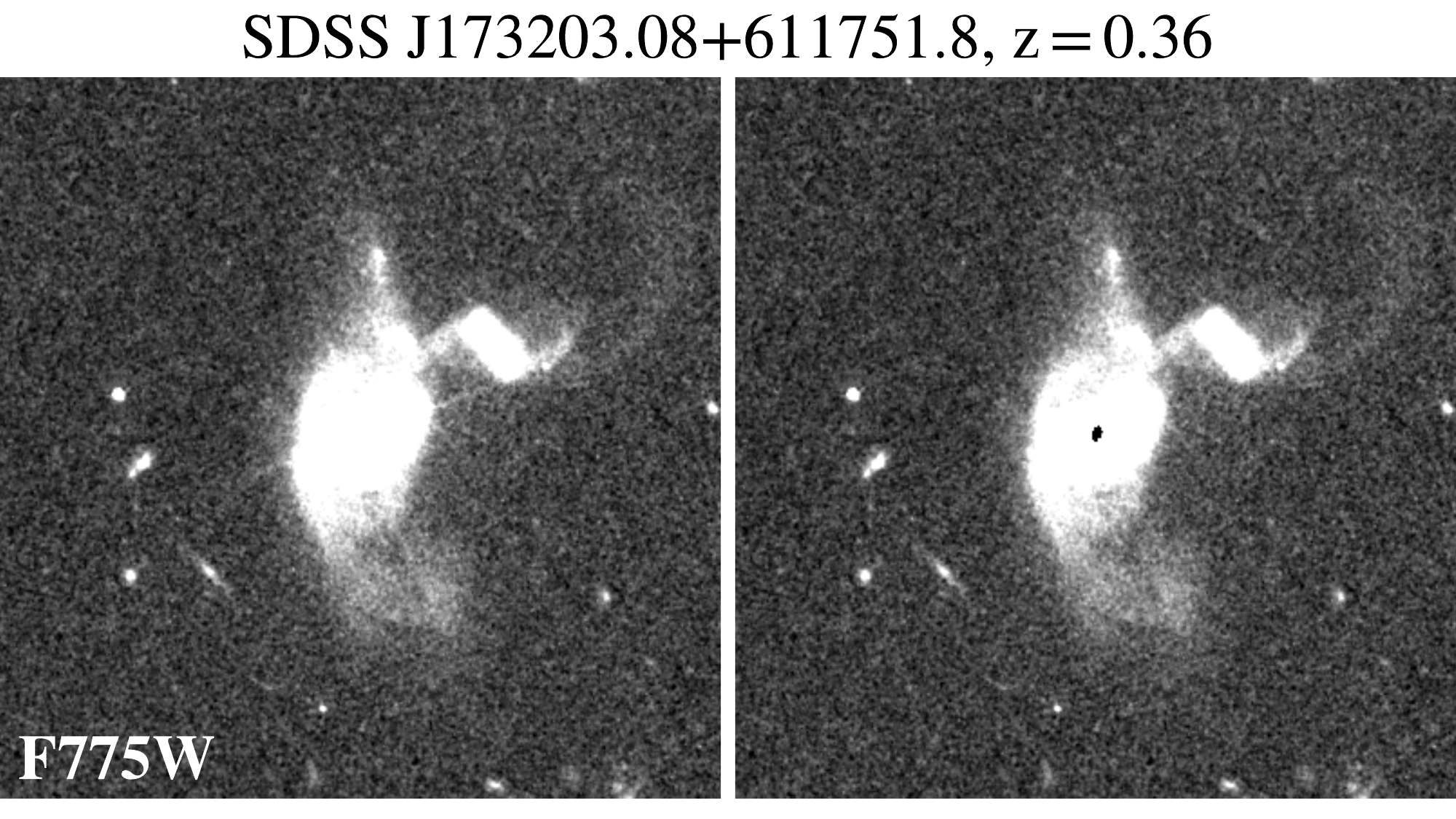}{0.45\textwidth}{}}
\caption{}
\end{figure*}
\renewcommand{\thefigure}{\arabic{figure}}



\end{document}